\documentclass[aip,reprint,floatfix,onecolumn,cha]{revtex4-1}
\usepackage{amsfonts,amssymb,amsmath,array}
\usepackage{comment}
\usepackage{cancel}
\usepackage{bm}
\usepackage[normalem]{ulem}

\usepackage{hyperref}
\hypersetup{colorlinks=true, linkcolor=black, citecolor=black, urlcolor=black}

\usepackage[final]{graphicx}
\usepackage{graphicx}
\usepackage{epstopdf}
\usepackage{mathtools}
\graphicspath{{./Figures/}} 

\usepackage[usenames]{color} 
\usepackage{xcolor}
\usepackage{soul} 

\begin{document}

\title{Emergent hydrodynamics of chiral active fluids: vortices, bubbles and odd diffusion}

\author{Umberto Marini Bettolo Marconi}
\affiliation{School of Sciences and Technology, University of Camerino, Via Madonna delle Carceri (Italy)}

\author{Alessandro Petrini}
\affiliation{Sapienza University of Rome,  P.le A. Moro 2, Rome (Italy).}

\author{Rapha\"el Maire}
\affiliation{Paris-Saclay University, CNRS, Laboratoire de Physique des Solides 91405 Orsay (France)}

\author{Lorenzo Caprini}
\affiliation{Sapienza University of Rome, P.le A. Moro 2, Rome (Italy)}
\email{lorenzo.caprini@uniroma1.it}

\newcommand{\bea}{\begin{eqnarray}}   
\newcommand{\eea}{\end{eqnarray}}

\newcommand{\beq}{\begin{equation}} 
\newcommand{\eeq}{\end{equation}}

\newcommand{\uu}{\boldsymbol{u}}
\newcommand{\br}{{\bf r}}
\newcommand{\bu}{{\bf u}}
\newcommand{\bv}{{\bf v}}
\newcommand{\bR}{{\bf R}}
\newcommand{\bx}{{ \bf x}}
\newcommand{\vv}{{\bf v}}
\newcommand{\bn}{{\bf n}}
\newcommand{\mb}{{\bf m}}
\newcommand{\bq}{{\bf q}}
\newcommand{\ff}{{\bf f}}
\newcommand{\bK}{{\bf K}}
\newcommand{\rb}{{\bar r}}
\newcommand{\rr}{{\bf r}}
\newcommand{\eb}{{\bf e}}
\newcommand{\ebh}{\hat{\bf e}}
\newcommand{\fvec}{{\bf F}}

\newcommand{\bnabvn}{{\bfnabla_{{\bf v}_i}}}
\newcommand{\bnabrn}{{\bfnabla_{{\bf r}_i}}}
\newcommand{\bnabri}{{\bfnabla_{{\bf r}_i}}}
\newcommand{\bnabr}{{\bfnabla_{{\bf r}}}}
\newcommand{\bnabv}{{\bfnabla_{{\bf v}}}}
\newcommand{\bnabeta}{{\bfnabla_{{}}}}
\newcommand{\bfnabla}{\mbox{\boldmath $\nabla$}}
\newcommand{\bnabvin}{{\bfnabla_{{\bf v}_i}}}

\newcommand{\unittheta}{\boldsymbol{\hat{\theta}}}
\newcommand{\bfxi}{\boldsymbol{\xi}}
 \newcommand{\bepsilon}{\boldsymbol{\epsilon}}
\newcommand{\kk}{\boldsymbol{\kappa}}
\newcommand{\greeketabold}{\boldsymbol{\eta}}
\newcommand{\xxi}{\boldsymbol{\xi}}
\newcommand{\cchi}{\boldsymbol{\chi}}
\newcommand{\bomega}{\boldsymbol{\Omega}}
\newcommand{\smallomega}{\boldsymbol{\omega}}
\date{\today}

\newcommand{\comm}[1]{\textcolor{red}{ #1}}
\newcommand{\edit}[1]{\textcolor{blue}{ #1}}

\begin{abstract}
Starting from a microscopic multiparticle Langevin equation, we systematically derive a hydrodynamic description in terms of density and momentum fields for chiral active particles interacting via standard repulsive and nonlocal odd forces.
These odd interactions are reciprocal but non-conservative: they are non-potential forces, as they act perpendicular to the vector joining any pair of particles. As a result, the torques that two particles exert on one another are non-reciprocal. The ensuing macroscopic continuum description consists of a continuity equation for the density and a generalized compressible Navier-Stokes equation for the fluid velocity. The latter includes a chirality-induced torque density term and an odd viscosity contribution.
Our theory predicts the emergence of odd diffusivity, edge currents, and an inhomogeneous phase -- characterized by bubble-like structures -- recently observed in simulations. Specifically, the theory exhibits a linear instability arising from the interplay between odd viscosity and torque density, and admits steady-state inhomogeneous solutions featuring bubbles and vortices, in agreement with numerical simulations. Our findings can be tested experimentally in systems of granular spinners or rotating microorganisms suspended in a fluid.
\end{abstract}

\date{\today}

\maketitle


\section{Introduction}

Active materials~\cite{ramaswamy2010mechanics,marchetti2013hydrodynamics,elgeti2015physics,bechinger2016active} have recently attracted significant attention due to their distinctive properties, representing a new class of nonequilibrium systems currently under intense investigation.
In these systems, the continuous injection of energy at the microscale drives the motion of individual constituents and breaks time-reversal symmetry~\cite{fodor2022irreversibility}. In many cases, an intrinsic degree of chirality leads to broken parity at the microscopic level, inducing circular or helical motion~\cite{lowen2016chirality,liebchen2022chiral}.

Chiral motion is typically generated by external torques acting on the particles, which may originate from different physical mechanisms. For instance, active colloids subjected to external magnetic fields perpendicular to the plane of motion exhibit chiral behavior and spinning dynamics~\cite{soni2019odd,mecke2023simultaneous}.
More generally, microswimmers with rotationally asymmetric propulsion mechanisms follow curved or rotating trajectories. This is the case for sperm cells~\cite{woolley2003motility}, which display chiral swimming paths due to asymmetric flagellar beating, as well as bacteria moving near surfaces~\cite{petroff2015fast}. Similar circular behavior is observed in L-shaped colloids~\cite{kummel2013circular} and in Janus colloidal beads with catalytic patches that are positioned at a fixed angle relative to one another~\cite{ebbens2010self}.
Chiral dynamics also arise in more complex biological systems, including cells~\cite{xu2007polarity}, algae~\cite{huang2021circular}, droplets~\cite{carenza2019rotation}, and starfish embryos~\cite{tan2022odd}. Moreover, chiral behavior can be reproduced at the macroscopic scale by suitably designing chiral active granular particles~\cite{siebers2023exploiting}. Examples include chiral rotors spinning due to airflow~\cite{lopez2022chirality}, rotationally asymmetric granular particles driven by internal motors~\cite{carrillo2025depinning}, or particles actuated by a vibrating plate~\cite{scholz2018rotating,caprini2025spontaneous}.

To reproduce the behavior of these systems, chirality is often incorporated in the dynamics through a constant angular velocity~\cite{lowen2016chirality}, which drives active Brownian particles along circular trajectories, both in free space~\cite{van2008dynamics,kummel2013circular} and under confinement~\cite{caprini2023chiral}.
This circular driving reduces both the effective speed and the diffusivity of active particles~\cite{van2008dynamics,sevilla2016diffusion,caprini2019active}, thereby altering the melting point~\cite{kuroda2025long} and suppressing clustering and motility-induced phase separation (MIPS)~\cite{liao2018clustering,ma2022dynamical,bickmann2022analytical} (see Ref.~\onlinecite{cates2015motility} for a review of MIPS).
More generally, circular driving alone can induce a wide range of collective phenomena in glasses~\cite{debets2023glassy}, liquids~\cite{kuroda2023microscopic}, and crystals~\cite{shee2024emergent,marconi2025spontaneous}. These include the formation of vortices~\cite{zhang2020reconfigurable,kruk2020traveling,liao2021emergent} and self-reverting vorticity~\cite{caprini2024self} in attractive systems. In the presence of alignment mechanisms, additional collective states emerge~\cite{kreienkamp2022clustering}, such as microflock patterns~\cite{liebchen2017collective}, traveling waves~\cite{liebchen2016pattern}, and self-rotating crystallites~\cite{huang2020dynamical,musacchio2025circling}.

When two or more chiral particles interact, they exert forces that break the parity symmetry and have no analogue in non-chiral systems. Indeed, these effective forces, termed odd interactions~\cite{caprini2025bubble,caprini2025modeling}, act transversely with respect to standard potential-derived interactions and are therefore non-conservative. Odd interactions can be generated by rotational friction in colliding granular spinners, or by hydrodynamic interactions when two swimmers rotate in a fluid, as in the case of algae~\cite{drescher2009dancing} or starfish embryos~\cite{tan2022odd}.
These forces underlie striking collective behaviors, such as cluster rotation observed in starfish colonies~\cite{tan2022odd}, as well as symmetry-protected surface currents in the vicinity of walls or at the edge of a cluster, as reported in Ref.~\onlinecite{caporusso2024phase} in the presence of attractive interactions.
In addition, odd interactions induce spatial velocity correlations~\cite{caprini2025odd} in crystalline phases and can drive a chirality-induced phase transition from a homogeneous state to an inhomogeneous phase characterized by bubbles, i.e., empty, approximately circular regions in a liquid. This phase, termed BIO (bubbles induced by odd interactions), was discovered in Ref.~\onlinecite{caprini2025bubble} and represents a general collective behavior uniquely arising from chirality.
A similar phenomenon was observed by Di Gregorio et al.~\cite{digregorio2025phase} using a different model for granular spinners subject to tangential friction, inspired by earlier studies on granular systems~\cite{brilliantov1996model}.
Finally, we note that the BIO phase is reminiscent of cavitation phenomena induced by a rotating particle coupled to a viscous fluid through hydrodynamic interactions, as simulated using lattice Boltzmann techniques~\cite{shen2023collective}.

Chiral systems are often modeled at a coarse-grained level using elastodynamic or hydrodynamic theories~\cite{fruchart2023odd}. Owing to broken parity symmetry, these theories differ fundamentally from their equilibrium counterparts and are characterized by odd elasticity~\cite{scheibner2020odd, alexander2021layered, braverman2021topological,shankar2024active,kole2024chirality,lee2025odd} in solids and odd viscosity~\cite{banerjee2017odd, markovich2021odd, reichhardt2022active, lou2022odd, hosaka2023lorentz, machado2023hamiltonian, markovich2021odd,markovich2025chiral} -- also known as Hall viscosity -- in chiral fluids.
Odd elasticity is a non-energy-conserving property found in certain solid-like systems, such as mechanical metamaterials, and is associated with antisymmetric shear and bulk elastic moduli. It underlies unconventional mechanical responses, including lateral deformation under compression and wave propagation, even in the overdamped regime.
The concept of odd viscosity was introduced into hydrodynamics by Avron~\cite{avron1998odd} and has recently been observed experimentally in both electron fluids~\cite{berdyugin2019measuring} and active-matter systems~\cite{soni2018free}. Unlike ordinary (even) viscosity, which is dissipative, odd viscosity~\cite{han2021fluctuating,reichhardt2022active,hosaka2023lorentz,markovich2024nonreciprocity} is reactive and does not lead to energy dissipation. Ordinary shear viscosity, being proportional to the strain rate, dissipates energy through friction between adjacent fluid layers moving at different velocities, while bulk viscosity characterizes a fluid’s resistance to compression or expansion and is likewise dissipative. In contrast, odd viscosity conserves energy and, in the case of shear flow, gives rise to a stress component perpendicular to the flow direction~\cite{lapa2014swimming,deshpande2024odd}.

While the odd elasticity tensor can be derived directly from microscopic dynamics governed by odd interactions~\cite{caprini2025odd}, the form of the odd viscosity tensor is typically introduced phenomenologically to capture the parity-breaking induced by chirality. We follow this phenomenological approach here, despite the recent proposal of a microscopic derivation of odd viscosity in Ref.~\onlinecite{eren2025collisional}.
Crucially, however, a direct connection between hydrodynamic descriptions and the emergent collective behavior -- specifically, the BIO phase -- numerically observed in chiral systems remains an open problem.

\begin{figure}
\includegraphics[width=\linewidth,clip=true]{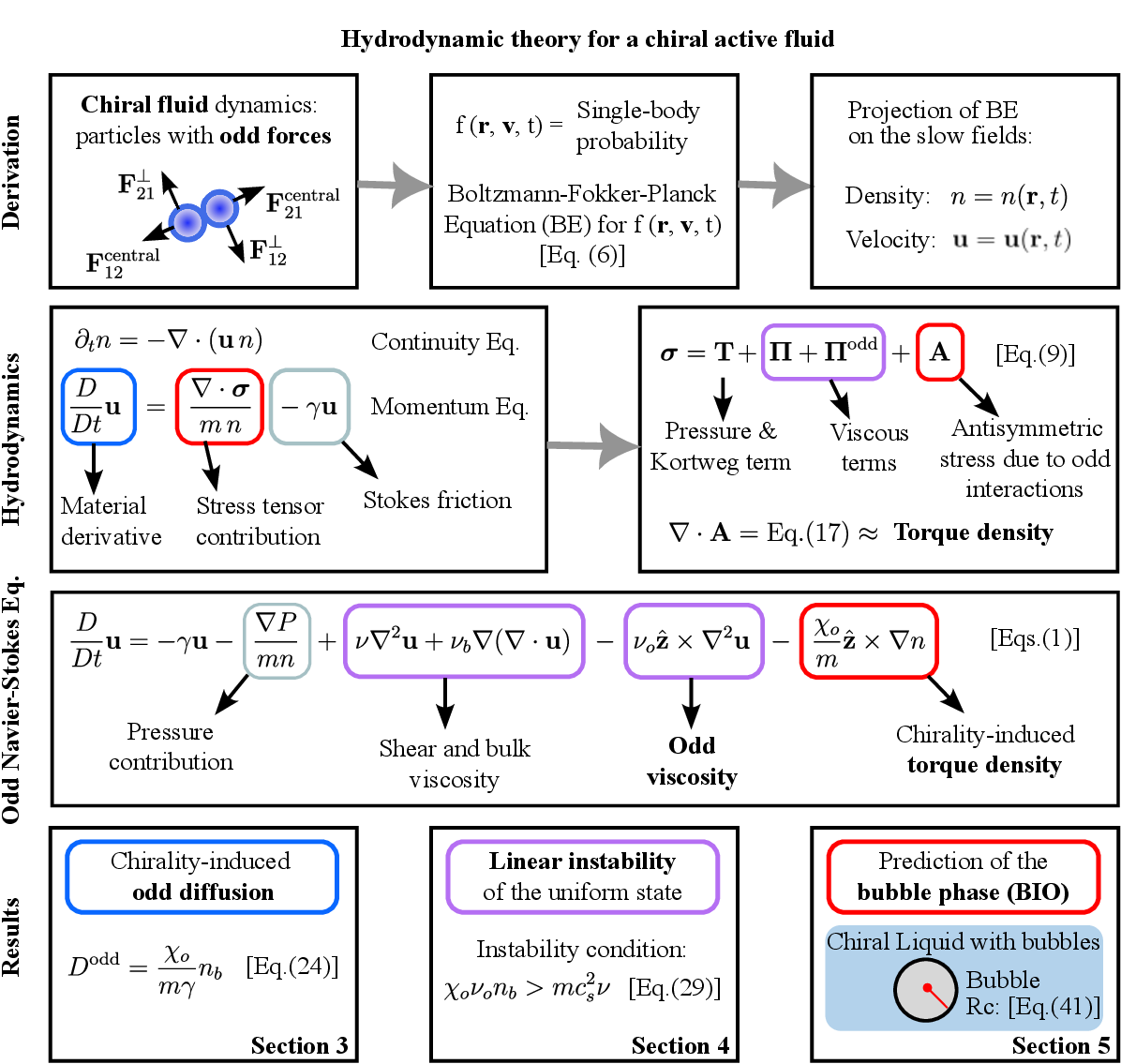}
\caption{ {\textbf{Hydrodynamics of a chiral active fluid}}.
The panels of the figure illustrate the main logical steps of this work and highlight the key results. Starting from a particle-based model of a chiral active fluid subject to odd interactions, we derive the Fokker–Planck equation for the $N$-particle distribution and close the BBGKY hierarchy at the single-particle level, obtaining a Boltzmann-like equation for the single-body distribution. 
Projecting onto the subspace spanned by the density $n = n(\mathbf{r}, t)$ and the velocity field $\mathbf{u} = \mathbf{u}(\mathbf{r}, t)$, we obtain a hydrodynamic description. The resulting chiral active fluid is governed by a compressible odd Navier–Stokes equation, featuring odd viscosity and a chirality-induced torque density. Our theory predicts that (i) odd interactions generate transverse diffusion; (ii) the interplay between odd viscosity and torque density destabilizes the homogeneous state; and (iii) this instability gives rise to an inhomogeneous phase characterized by bubbles, termed BIO \cite{caprini2025bubble}.}
\label{fig:Fig0}
\end{figure}

Here, starting from a microscopic Langevin description of non-motile chiral active particles, we derive macroscopic continuum equations for the density and fluid velocity governing a chiral active fluid. The resulting hydrodynamic description differs from the continuity and Navier–Stokes equations of an ordinary fluid by the presence of an additional effective odd force term proportional to the skew density gradient, which can be identified as a chirality-induced torque density (Fig.~\ref{fig:Fig0}).
We discover that this torque density generates odd diffusion and, when combined with odd viscosity, a linear instability of the homogeneous phase with constant density and vanishing momentum. The linear stability analysis leads to a phase diagram in qualitative agreement with previous results based on molecular dynamics simulations. In addition, the theory admits a steady-state solution in the inviscid limit featuring cavities and vorticity, which is reminiscent of the BIO phase. 

We begin by introducing the model and deriving the coarse-grained hydrodynamic theory in Sec.~\ref{sec:hydro}. The consequences of this description—namely odd diffusion and the linear stability of the homogeneous phase—are examined in Secs.~\ref{sec:OddDiffusion} and~\ref{sec:linearStability}. In Sec.~\ref{sec:nonlinear}, we present an approximate solution of the nonlinear equations, before concluding with a summary and outlook.

\section{Hydrodynamic theory for chiral active particles}\label{sec:hydro}

In this section, we address the classical problem of statistical mechanics: deriving the laws that govern the large-scale behavior of a many-particle system starting from the dynamics of its elementary constituents. For equilibrium systems, this procedure is, in principle, straightforward and leads to the Navier–Stokes equation for a compressible fluid.

By contrast, in chiral active matter governed by odd (transverse) interactions -- which break both detailed balance and parity symmetry -- the same protocol yields a hydrodynamic theory containing additional terms beyond those of an equilibrium fluid. Specifically, we show that a chiral system is described by a continuity equation for the number density $n(\mathbf{r},t)$ and a modified Navier–Stokes equation for the velocity field $\mathbf{u}(\mathbf{r},t)$ (see below for precise field definitions), with the following form (see Fig.~\ref{fig:Fig0}):
\begin{subequations}
\bea
&&\partial_{t}n(\rr,t)= - \boldsymbol{\nabla}\cdot(n(\rr,t)\bu(\rr,t))
 \label{eq:continuity}\\
&&
\partial_{t} \bu + \bu\cdot\boldsymbol{\nabla}\bu  = -\frac{1}{ m n}\Bigl( \bfnabla P  -\kappa n \bfnabla \bfnabla^2 n\Bigr)
 - \gamma \bu +
\frac{\eta}{mn} \bfnabla^2 \bu +\frac{\zeta}{mn} \bfnabla(\bfnabla\cdot\bu)
 - \frac{\eta_o}{mn} {\bf \hat z}\times\bfnabla^2 \bu -\dfrac{\chi_{o}}{m}  {\bf \hat z}\times \bfnabla n  \,\rm{.}
 \label{NSeq}
   \eea
 \end{subequations}
Here, $m$ represents the particle mass, $P$ is a standard pressure contribution due to thermal agitation and repulsive interactions, while the term $-\kappa n \bfnabla \bfnabla^2 n$ directly follows from the Korteweg stress tensor as in an equilibrium fluid.
Our system is additionally subject to a friction force $-\gamma \mathbf{u}$ accounting for the dry friction with a substrate and to standard viscous terms proportional to the velocity gradients (third and fourth terms on the right-hand side of Eq.~\eqref{NSeq}), through the shear viscosity $\eta$ and the bulk viscosity $\zeta$.

Beyond conventional terms, the hydrodynamics of chiral active systems is characterized by two additional contributions: i) an odd viscous term $\propto \eta_o {\bf \hat z}\times\bfnabla^2 \bu$, proportional to the odd viscosity $\eta_o$, which mixes the Cartesian components of $\boldsymbol{\nabla}^2 \mathbf{u}$ through the cross product with a unit vector $\bf \hat z$ such that $\hat{\mathbf z}\times \bfnabla^2 \mathbf u = \big(-\bfnabla^2 u_y,\bfnabla^2 u_x\big)$ (fifth term on the right-hand side of Eq.~\eqref{NSeq}); ii) an extra contribution, $\chi_o{\bf \hat z}\times \bfnabla n$ (sixth term on the right-hand side of Eq.~\eqref{NSeq}) with strength $\chi_o$ which mixes the Cartesian component of $\bfnabla n$.
This additional term arises from a chirality-induced torque density 
\begin{equation}
    \chi(\rr,t)=\dfrac{1}{2}\chi_o n^2(\rr,t) \,\rm{,}
    \label{torquedensitydef}
\end{equation}
appearing in the Navier-Stokes equation~\eqref{NSeq} as $-{\bf \hat z}\times \bfnabla \chi/(mn)$.
While the odd viscous term (i) has been extensively investigated through numerical and analytical studies, the presence of a chirality-induced torque density (term (ii) in Eq.~\eqref{NSeq}) constitutes our first result. Compared to conventional and odd viscous contributions, this torque density is peculiar for two reasons. First, it originates from a contribution to the stress tensor that explicitly breaks parity symmetry in the momentum equation, being antisymmetric under the exchange of its Cartesian components. Second, unlike viscous terms, it is not proportional to velocity gradients but instead depends on the density, similarly to a pressure tensor, while containing only antisymmetric off-diagonal elements. As a consequence, this term corresponds to a force per unit volume that is tangential to the surface, whereas the divergence of the pressure tensor generates a force normal to the surface.

Although the present paper focuses on a two-dimensional system, we adopt the conventional three-dimensional terminology throughout. Accordingly, we use the term “pressure” to denote a line tension, “volume” to denote an area, “surface” to denote a line, and so forth. For example, the curl operator has only the component normal to the plane in which the particles move. We believe that this choice of terminology improves the readability of the paper. In the remainder of this section, we present the derivation of the macroscopic hydrodynamic equations and the associated stress tensor, starting from the Langevin dynamics of chiral fluids interacting through odd (transverse) forces.

\subsection{Model for interacting chiral active particles}

The equations of motion for non-motile chiral active particles of mass $m$, interacting via repulsive and odd forces, are governed by underdamped dynamics for the particle positions, $\bm{r}_{i}$, and velocities, $\bm{v}_{i}$, which are given by:
\begin{equation}
m \dot \vv_i(t)=
\sum_{j\neq i} \big({\bf F}_{ij}^{Central}+{\bf F}_{ij}^{\perp}\big)-m\gamma  \vv_i(t)
+ \sqrt{2m\gamma T} \,\xxi_i(t)  \,\rm{,}
 \label{Langevin1}
\end{equation}
where $\gamma$ is the friction coefficient and $T$ the temperature of the solvent, assuming a unit Boltzmann constant $k_B=1$. Thus, particles experience a Stokes friction force $-m\gamma \vv$ with the medium, through which they exchange energy due to a random force $\sqrt{2m\gamma T} \,\xxi_i(t)$.  
Here, the term $\xxi_i$ corresponds to a white noise vector with unit variance and zero average, which accounts for the random collision of the environmental molecules. 
Particles $i$ and $j$ interact through two force terms, ${\bf F}_{ij}^{Central}$ and ${\bf F}_{ij}^{\perp}$, depending on the particle coordinates.
The first term ${\bf F}_{ij}^{Central}=-\bnabrn U^{WCA}(|\rr_i-\rr_j|)$ is a standard pair force, derived from a potential, for instance, a pure repulsive Weeks-Chandler-Andersen potential $U^{WCA}(|\rr_i-\rr_j|)$, i.e., a Lennard-Jones potential cut and shifted to zero at the cutoff. 
In what follows, this choice is not fundamental even if, for simplicity, we assume that the potential is short-range.

The second force term ${\bf F}_{ij}^{\perp}$ corresponds to the odd interactions effectively generated by chirality and acts perpendicularly to the relative distance between the particle's centers. To keep this structure, ${\bf F}_{ij}^{\perp}$ can be expressed as 
\beq
\fvec_{ij}^{\perp}= - \bfnabla_i U^\perp(|\rr_i-\rr_j|) \times \hat{\bf z}\,\rm{,}
\label{pseudopotentialforce}
\eeq
where $U^\perp(\rr_i-\rr_j)$ is a scalar function which depends on the separation of the particles $i$ and $j$, and $\hat{\bf z}$ is a vector orthogonal to the plane of motion
\footnote{The interaction~\eqref{pseudopotentialforce} belongs to the class of curl forces, forces whose curl does not vanish (see M. Berry and P. Shukla, Journal of Physics A: Mathematical and Theoretical 45, 305201 (2012)).}.

As discussed in previous work~\cite{caprini2025bubble,marconi2025spontaneous}, the force~\eqref{pseudopotentialforce} is translationally invariant, yet non-conservative since it cannot be expressed as the gradient of a potential. In particular, linear momentum is conserved because $\fvec^\perp_{ji}=-\fvec^\perp_{ij}$. However, explicit parity breaking prevents the torques that particles exert on one another from canceling, so angular momentum and energy are not conserved. Thus, although the forces are reciprocal because they satisfy Newton's third law, the torques are non-reciprocal.
The non-reciprocity of these interactions is not a property of any fundamental force, such as electromagnetic or gravitational interactions, but it appears at a mesoscopic level as a result of collisions subject to rotational friction or mediated by a fluid, as mentioned in the Introduction. Hence, the transverse force, $\fvec_{ij}^{\perp}$, must be viewed as the result of a coarse graining procedure which traces out some internal degrees of freedom, such as rotations of the particles around their axes (see Appendix~\ref{heuristicappendix}) and allows a description only in terms of the translational degrees of freedom of the particles.
The form of the force~\eqref{pseudopotentialforce} is very convenient because it is simple to implement numerically and greatly simplifies the analytical calculations with respect to other models where the transverse interaction depends on the velocities of the particles, as in Ref.~\onlinecite{digregorio2025phase}  or even on the hydrodynamic fields in the case of the Lattice Boltzmann simulations~\cite{shen2023collective}.

From the Langevin dynamics~\eqref{Langevin1}, one can derive the Kramers-Fokker-Planck equation describing the evolution of the multidimensional phase space density distribution function, $f_N=f_N(\{\rr,\vv\},t)$, where $\{\rr,\vv\}$ indicates the position and velocity coordinates of the $N$ particles~\cite{Risken}.
This equation reads
\begin{eqnarray}
&&
\Bigl(\frac{\partial}{\partial t}+\sum_i \Bigl[\vv_i\cdot\bnabrn  \Bigr]
\Bigr) f_N 
- \sum_i \gamma \Bigr[\bnabvin\cdot (\vv_i f_N )+v_T^2\bnabvin^2f_N  \Bigr] 
= - \frac{1}{m} \sum_i \sum_{j \neq i}\Bigl[ 
\mathbf{F}_{ij}^{Central} + \mathbf{F}_{ij}^{\perp} \Bigr]
\cdot\bnabvin f_N \, \rm{,}
\label{many4}
\end{eqnarray}
where $v_T^2=\frac{T}{m}$ is the thermal velocity of the particles.
This equation will be the starting point to derive the hydrodynamic equations.

\subsection{Hydrodynamic theory for chiral particles subject to transverse forces}

To proceed further, we introduce the reduced single-particle distribution function, $f=f(\rr,\vv,t)$, i.e.\ the probability of finding a particle at time $t$ with position $\rr$ and velocity $\vv$. The distribution $f=f(\rr,\vv,t)$ can be obtained by integrating the full probability distribution $f_N$ over the coordinates (positions and velocities) of $N-1$ particles:
$f(\rr,\vv,t)= \Pi_{i=2}^N\int  d\vv_i d \rr_i  f_N(\{\rr,\vv\},t)$. 
 By integrating the Kramers-Fokker-Planck equation (Eq.~\eqref{many4}) over the $N-1$ particle coordinates, we obtain the Fokker-Planck-Boltzmann equation describing the temporal evolution of $f$:
\bea
&&
\Bigl(\partial_t+\vv\cdot\bnabr \Bigr) f  - \gamma \Bigl[\bnabv\cdot (\vv f) +v_T^2\boldsymbol{\nabla}^2_{\mathbf{v}}f \Bigr] 
= \frac{1}{m}  \int d\rr' \int d\vv'\Bigl[ \bnabr U^{WCA}(|\rr-\rr'|) + \bnabr U^{\perp}(|\rr-\rr'|) \times {\bf \hat z} \Bigr] \cdot\bnabv  f_2\,\rm{.}
\label{FPBE1}
\eea
Equation~\eqref{FPBE1} for the single-body probability distribution $f(\rr,\vv,t)$ is linear, but is not closed due to the presence of the two-body probability distribution $f_2=f_2(\rr, \vv, \rr', \vv', t)$. The function $f_2$ is defined as the integral of $f_N$ over $N-2$ particle coordinates and represents the probability of finding at time $t$ two particles, one at coordinates $(\rr, \vv)$ and the other at coordinates $(\rr', \vv')$. 
We shall not try to go beyond the first level of the Bogoliubov–Born–Green–Kirkwood–Yvon (BBGKY) hierarchy~\cite{hansen2013theory} and adopt a very simple mean-field ansatz to close Eq.~\eqref{FPBE1}. The strategy amounts to factorizing $f_2$ into the product of two single-particle distribution functions and the pair correlation function.

We define the number density $n(\rr,t)$ and the fluid velocity $\bu(\rr,t)$ at time $t$ and position $\rr$, as the zero and first moment of the distribution $f$. By integrating in the velocity space, we obtain
\beq
\begin{pmatrix} n(\rr,t) \\  n(\rr,t)\bu(\rr,t) \end{pmatrix}\equiv
\int d\vv \, f(\rr,\vv,t) 
\begin{pmatrix} 1 \\ \vv  \end{pmatrix} \,\rm{.}
\label{hydrofields}
\eeq
To derive hydrodynamic equations for $n(\rr,t)$ and $\bu(\rr,t)$, we multiply Eq.~\eqref{FPBE1} by $1$ and $m\vv$, respectively, and integrate with respect to the velocity.
The resulting equations for the  hydrodynamic fields are:
\begin{subequations}
\label{eq:hydro_all}
\begin{eqnarray}&&
\partial_{t}n(\rr,t)= - \boldsymbol{\nabla}\cdot(n(\rr,t)\bu(\rr,t))
 \label{continuity1}
\\&&
\partial_{t}\bu(\rr,t) + \bu\cdot\boldsymbol{\nabla}\bu =  \frac{1}{m n}\boldsymbol{\nabla}\cdot\boldsymbol{\sigma} - \gamma \bu (\rr,t) \,\rm{.}
\label{velocityeq}
\end{eqnarray}
\end{subequations}
Equations~\eqref{continuity1} and~\eqref{velocityeq} represent the continuity equation and the Navier-Stokes equation of the model in the presence of friction, central, and transverse forces.
The symbol $\boldsymbol{\sigma} $ denotes the stress tensor, which can be decomposed into a kinetic energy $\sigma^{(kin)}$ and a potential part $\sigma^{(pot)}$ due to conservative interactions, as in equilibrium, and an additional contribution directly arising from chirality through odd interactions:
\bea&&
\partial_\beta\sigma_{\alpha\beta}=\partial_\beta\sigma^{(kin)}_{\alpha\beta}+\partial_\beta\sigma^{(pot)}_{\alpha\beta}
+\partial_\beta A_{\alpha\beta} = \partial_{\beta}\sigma^{(sym)}_{\alpha\beta} + \partial_\beta A_{\alpha\beta}\,\rm{.}
\label{sigmatotal}
\eea
The kinetic part $\sigma^{(kin)}$ is defined as in equilibrium and is directly related to the temperature field, while the potential part $\sigma^{(pot)}$ is generated by the interaction potential and thus involves $f_2$:
\bea
&&
\partial_\beta\sigma^{(kin)}_{\alpha\beta}(\rr,t)=-m\partial_\beta \int d\vv  (\vv-\bu)_\alpha(\vv-\bu)_\beta f(\rr,\vv,t) 
\label{kineticsigma}\\&&
\partial_\beta  \sigma_{\alpha\beta}^{(pot)}(\rr,t)=\int d\vv    v_\alpha\,  {\bfnabla_{\bf v}} \cdot   \int d\rr'\int d\vv' 
f_2(\rr,\vv; \rr', \vv',t){\bfnabla_{\bf r}}U^{WCA}(|\rr-\rr'|) \,\rm{.}
\label{potentialsigma}
\eea
Here, the term~\eqref{kineticsigma} is the low-density contribution which prevails in the dilute regime, whereas~\eqref{potentialsigma} originates from the interactions associated with the interparticle potential $U^{WCA}$.
While the above terms characterize the hydrodynamic theory of an equilibrium fluid, odd interactions generate the following additional term
\bea
\partial_\beta A_{\alpha\beta}(\rr,t)&&= \int d\vv \, v_\alpha \int d\rr' \int d\vv'  \Bigl[ \bnabr U^{\perp}(|\rr-\rr'|) \times {\bf \hat z} \Bigr]\cdot\bnabv
f_2(\rr,\vv,\rr',\vv',t) \,\rm{,}
\eea
which again involves the two-body probability distribution $f_2$ and needs a suitable closure. We notice that $\bf A$ is the only antisymmetric component and therefore it is responsible for the production of angular momentum in the system. 
In the following sections, we will group the kinetic (Eq.~\eqref{kineticsigma}) and potential (Eq.~\eqref{potentialsigma}) contributions to the symmetric stress tensor into $\boldsymbol{\sigma}^{(sym)}$, thereby decomposing the stress tensor as in the last equality of Eq.~\eqref{sigmatotal}. 

\subsection{Closed expression for the Stress tensor} 

The exact expression of the terms of the second rank tensor $\boldsymbol{\sigma}^{(sym)}$ indicated in Eqs.~\eqref{kineticsigma} and~\eqref{potentialsigma} is not known, and it cannot be obtained from kinetics, apart from the limiting case of rarefied gases.
It is useful to further decompose the symmetric stress as
\beq
\sigma_{\alpha\beta}^{(sym)}= T_{\alpha\beta}+\Pi_{\alpha\beta}+\Pi_{\alpha\beta}^{odd}\,\rm{,}
\label{pressuretensor}
\eeq
where $T_{\alpha\beta}$ denotes the non-dissipative (reactive) contribution, $\Pi_{\alpha\beta}$ is the conventional viscous stress tensor, and $\Pi_{\alpha\beta}^{\mathrm{odd}}$ is the odd viscous stress tensor associated with odd viscosity in chiral fluids.

The three terms in Eq.~\eqref{pressuretensor} can be approximated in terms of density $n$ and velocity field $\mathbf{u}$. 
The first one $T_{\alpha\beta}$ contains the contributions from the pressure, the van der Waals surface term, and the Korteweg capillary stress:
 \beq
 T_{\alpha\beta}=\Bigl( -P(n)+\frac{\kappa}{2}(\bfnabla n)^2+\kappa n\bfnabla^2 n   \Bigr)\delta_{\alpha\beta}
 -\kappa (\partial_\alpha n)   (\partial_\beta n) \,\rm{.}
 \eeq
Here, the pressure $P$ depends on the number density $n$ and is given by the equation of state of the fluid. The remaining contributions in the expression for $ T_{\alpha\beta}$ represent the Korteweg stress tensor. These terms account for the interfacial contribution to the total stress stemming from density gradients, with $\kappa$ the interfacial stiffness coefficient~\cite{anderson1998diffuse}.

The viscous tensor, $\Pi_{\alpha\beta}$, can be written as a rotationally invariant linear combination of velocity field gradients, 
\beq
\Pi_{\alpha\beta}=
\eta\left(\frac{\partial u_\alpha}{\partial x_\beta}+\frac{\partial u_\beta}{\partial x_\alpha}\right) +(\zeta-\eta )\nabla\cdot \bu \,\delta_{\alpha\beta} \,\rm{,}
\label{viscoustensorsym}
\eeq
where $\eta$ denotes the dynamical shear viscosity coefficient and $\zeta$ the bulk viscosity. These terms are functions of the density and, as usual in conventional fluids, the ratios $\nu=\eta/(mn)$ and $\nu_b=\zeta/(mn)$ are assumed to be constant according to the so-called phenomenological constitutive relations~\cite{epstein2020time}, which define the shear and bulk kinetic viscosity, respectively.
Therefore, the first two terms of Eq.~\eqref{pressuretensor} describe only the standard contributions to the stress tensor arising in simple fluids.
Instead, the odd viscous stress tensor $\Pi^{odd}_{\alpha\beta}$ is discussed in the next section. 
In the spirit of the present treatment, we shall not try to derive expressions for $P$, $\eta$, and $\zeta$ from our microscopic model, a task that we leave to future work.

\subsection{Closed expression for the Stress tensor: the odd viscous term and the antisymmetric contribution} 

While in simple fluids the hydrodynamics is usually reproduced by a rotationally invariant symmetric stress tensor, in odd fluids, time reversal and parity are often broken. In this case, Avron~\cite{avron1998odd} has shown that the viscous tensor can have an extra contribution, ${\boldsymbol \Pi}^{odd} $,  the so-called odd viscous stress tensor (see Appendix~\ref{oddviscoustensor}), which it is expressed as a linear combination of velocity gradients and involves a new transport coefficient, e.g. the odd viscosity coefficient $\eta_o$. 
This term enters the Navier-Stokes equation~\eqref{velocityeq} in the following way:
\bea&&
\bfnabla \cdot  {\boldsymbol \Pi}^{odd} =-\eta_o  {\bf \hat z}\times \nabla^2  \bu \,\rm{,}
\label{viscousodd}
\eea
where $\nu_{o}=\eta_o/(nm)$ defines the kinematic odd viscosity.
We remark that ${\boldsymbol \Pi}^{odd} $ is symmetric with respect to the exchange of its two indices, is not dissipative, and represents a reactive term sometimes called odd pressure. It describes non-dissipative flows orthogonal to the perturbation directions. 

Finally, we derive a closed, simple expression for the anti-symmetric (odd-parity) stress, ${\bf A}$, in terms of the hydrodynamic fields for a homogeneous configuration.
To this purpose, we perform a Vlasov-like~\cite{spohn2012large} (mean field-like) approximation of the two particle distribution function ($ f_2\to f(\rr,\vv,t)   f(\rr',\vv',t)  $) and obtain
\bea
\partial_\beta A_{\alpha\beta}(\rr,t)&&= -\int d\vv \, v_\alpha \int d\rr' \int d\vv'  \Bigl[{\bf \hat z}\times \bnabr U^{\perp}(|\rr-\rr'|) \Bigr]\cdot\bnabv
f_2(\rr,\vv,\rr',\vv',t)\nonumber \\
&&
\approx -\int d\vv \sum_\beta v_\alpha \frac{\partial}{\partial v_\beta} f(\rr,\vv,t) \int d\rr' \int d\vv'  \Bigl[{\bf \hat z}\times \bnabr U^{\perp}(|\rr-\rr'|) \Bigr]_\beta
f(\rr',\vv',t)
\nonumber\\
&&
\approx 
n(\rr,t)\int d\rr' \, [{\bf \hat z}\times \bnabr U^{\perp}(|\rr-\rr'|)]_\alpha  n(\rr',t)=  n(\rr,t) f_\alpha^{\perp} (\rr,t) \,\rm{.}
\eea
The last equality defines the effective mean odd force ${\bf f}^{\perp}(\rr,t)$, generated by the presence of odd (transverse) interactions, in terms of a configurational integral.
This integral can be further simplified by expanding the density $n$ in gradients as illustrated in Appendix~\ref{effectiveoddforce}. By retaining only the first term in a density gradient expansion of ${\bf f}^{\perp}(\rr,t)$, we find:
\beq
 {\bf f}^{\perp}(\rr)=
-\chi_o\,  \hat{\bf z} \times \bfnabla n(\rr)\,\rm{.}
\label{perpendicularforce}
\eeq
Here, $\chi_o$ is a constant with dimensions $[\chi_o]=[mass \cdot length^4/time^2]$, which gives the strength of the transverse interaction, and reads
\beq
\chi_o= - \pi \int_0^\infty R^2 dR\,   \frac{ d U^{\perp}(R)}{\partial R}
\label{Aconstant}
\eeq
which is positive for a decreasing function of the particle distance $R$, such that $\frac{ d U^{\perp}(R)}{\partial R} <0$, as chosen in numerical simulations~\cite{caporusso2024phase, caprini2025bubble}.
Notice that $\mathbf{f}^{\perp}$ gives rise to large contributions in regions where the density is non-uniform. This leads to momentum currents at the boundary of a cluster or a bubble, as well as in the vicinity of a wall. By contrast, $\mathbf{f}^{\perp}$ remains small in the bulk of a homogeneous phase. The details of the procedure used to derive this term are reported in Appendix~\ref{effectiveoddforce}.

Let us remark that the tensor $\mathbf{A}$ can be written as $A_{\alpha\beta}=\epsilon_{\alpha\beta} \chi_{o} n^2/2
=\epsilon_{\alpha\beta} \,\chi(\rr,t)$, where $\chi$ is the torque density defined by Eq.~\eqref{torquedensitydef} and $\epsilon_{\alpha\beta}$ are the components of the rank-2 Levi–Civita tensor with $\epsilon_{xy}=-\epsilon_{yx}=1$ and $\epsilon_{xx}=\epsilon_{yy}=0$. This form highlights the distinction from the usual isotropic pressure term $P(\mathbf{r},t)\delta_{\alpha\beta}$. While the pressure quantifies the contact forces acting normal to a surface, $A_{\alpha\beta}$ characterizes an imbalance of forces acting tangentially to that surface. Spatial variations of $n$ render $\partial_\beta A_{\alpha\beta}=-\chi_o n(\hat{\mathbf{z}}\times\nabla n)_\alpha$ nonzero, yielding a transverse force density that can drive vorticity.
The antisymmetric tensor field ${\bf A}$ emerges from the collective behavior of a large number of microscopic degrees of freedom and is not a property of any individual particle. In other words, $A_{\alpha \beta}$ is a parity-odd stress and an emergent feature of a system with transverse interactions, a hallmark of a non-vanishing torque density. It manifests as a contact force and is a robust, measurable macroscopic property.
Unlike the viscous stress, $A_{\alpha \beta}$ is not proportional to velocity gradients; rather, it depends on the density, similar to the pressure tensor, and contains only antisymmetric elements. Furthermore, the divergence $\sum_\beta \nabla_\beta A_{\alpha \beta}$ corresponds to a force per unit volume (= area) tangential to a surface (curve) whose local normal is aligned with the $\alpha$ direction, whereas the divergence of the pressure tensor generates a force normal to the surface (curve).

Finally, using the expressions~\eqref{viscoustensorsym},~\eqref{viscousodd} and~\eqref{perpendicularforce} determining the stress contributions, we rewrite the hydrodynamic equation~\eqref{velocityeq} only in terms of the density and velocity fields:
\bea&&
\partial_{t} \bu + \bu\cdot\boldsymbol{\nabla}\bu  = -\dfrac{1}{ m n}\Bigl( \bfnabla P  -\kappa n \bfnabla \bfnabla^2 n\Bigr)
 - \gamma \bu +
\nu \bfnabla^2 \bu +\nu_b \bfnabla(\bfnabla\cdot\bu)
 - \nu_o \hat{\bf z}\times\bfnabla^2 \bu -\dfrac{\chi_o}{m}  \hat{\bf z}\times \bfnabla n  \,\rm{.}
 \label{velocita}
   \eea
 Equation~\eqref{velocita} to be solved together with Eq.~\eqref{continuity1} is of the form of a Navier-Stokes equation for a compressible fluid with friction, even and odd viscosity, and a parity-breaking term which induces a flux in a direction perpendicular to the density gradient 
  \footnote{
It is important not to confuse the antisymmetric stress tensor ${\bf A}$, a function of the density but not of the velocity field,  with the quantity which has the same physical dimensions and is often  called  odd pressure in the literature, and whose expression in the case of incompressible fluids is:
$$
P_o=\eta_o (\partial_x u_y-\partial_y u_x)=\eta_o \omega \, \rm{,}
$$
where in the last equality it is shown in terms of the local vorticity of the fluid $\omega=(\bfnabla\times \bu)_z$.
}.
In the following sections, we explore the hydrodynamic equations (Eq.~\eqref{velocita}) and derive the main phenomena governing a chiral active fluid.

\section{Chirality-induced odd diffusion}\label{sec:OddDiffusion}

We begin by evaluating the long-time, large-wavelength solution of the compressible Navier–Stokes equations for a chiral fluid. We find that odd interactions, mediated by chirality, generate odd diffusion. This means that the effective Fick’s law governing the density evolution involves a diffusion matrix $\mathbf{D}$ with antisymmetric off-diagonal elements, which can be identified as odd diffusion coefficients~\cite{hargus2021odd,kalz2022collisions,vega2022diffusive,kalz2024oscillatory}. These terms are typically associated with circular momentum currents in the steady state, typically induced by magnetic fields or single-particle rotating motion. Here, we demonstrate that odd diffusion can alternatively arise from odd (transverse) interactions, as predicted by our hydrodynamic theory.
 
By evaluating the momentum equation (Eq.~\eqref{velocita}) in the long-time limit, $t \gg 1/\gamma$, the convective acceleration and time derivative become negligible. In this regime, the momentum equation reduces to a local force balance and the particle current,
\begin{equation}
    \mathbf J(\mathbf r, t)=n(\mathbf r, t)\bu(\mathbf r, t)
\end{equation}
reduces to
\begin{eqnarray}
  \mathbf J(\mathbf r, t)=  \frac{1}{m \gamma }\bfnabla \cdot\boldsymbol{\sigma}^{(sym)}   + \frac{1}{m \gamma }{\bf f}^{\perp}(\rr,t) n(\rr,t) \,\rm{.}
\label{lasteqb}
\end{eqnarray}
Using the result~\eqref{perpendicularforce} and neglecting the viscous terms, i.e. approximating $\bfnabla\cdot \boldsymbol{\sigma}^{(sym)}\approx-\bfnabla P(\rr)  +\kappa n(\rr) \bfnabla \bfnabla^2 n(\rr)$, yields the following decomposition for the current:
\beq
\mathbf{J}= -\frac{1}{m\gamma}\bigl[\bfnabla P-\kappa n\bfnabla\bfnabla^2 n\bigr]-\frac{\chi_o}{m\gamma}n(\hat{\mathbf{z}}\times\bfnabla n)\,.
\eeq
The first bracket describes the conventional compressible diffusive flux: the pressure gradient and the interfacial stiffness term generate a current that tends to suppress density gradients and thus favors a uniform state with constant density. The second term is the odd contribution, which is proportional to the torque-density strength $\chi_o$ and acts transversely to$\bfnabla n$, generating a current perpendicular to the density gradient. As mentioned before, this behavior typically occurs whenever the system performs circular-like orbits~\cite{hargus2021odd}, as occurs in the case of odd interactions~\cite{caprini2025modeling}.

By linearizing and applying the Fourier transform (see Appendix~\ref{app:FourierAppendix} for the definition of the Fourier transform), we obtain an effective Fick's equation for the Fourier transform of the density current: $\hat J_\alpha(\mathbf{q},t)=-D_{\alpha\beta}(\mathbf{q})(iq_{\beta})\hat n(\mathbf{q},t)$ with a wavevector-dependent diffusion matrix:
\begin{equation}
\hspace{-2.5cm}\mathbf{D} = 
\left( \begin{array}{cc}
\dfrac{c_s^2}{\gamma} +\dfrac{\kappa n_b}{m\gamma} q^2 & -\dfrac{\chi_o}{m\gamma} n_b  \\ 
\dfrac{\chi_o}{m\gamma} n_b & \dfrac{c_s^2}{\gamma}+\dfrac{\kappa n_b}{m\gamma} q^2 \\ 
\end{array} \right) \,\rm{.}
\label{dxx}
\end{equation}
Here, $c_s \equiv \sqrt{(dP/dn)/m}\big|{n=n_b}$ denotes the sound speed of an underdamped equilibrium system at uniform density $n_b$. The diffusion matrix has two identical diagonal elements, reflecting the isotropy of the system, and these elements formally coincide with their equilibrium counterparts. In the long-wavelength limit, $\mathbf q \to 0$, the long-time diffusion coefficient reduces to $c_s^2/\gamma$. In the ideal-gas limit, this further simplifies to $D_{xx}=D_{yy}=T/(\gamma m)$, recovering the familiar Einstein relation between diffusivity, temperature, and friction.
By contrast, chirality generates an antisymmetric off-diagonal component, $D_{xy}=-D_{yx}=-\chi_o n_b/(m\gamma)$. Importantly, this antisymmetric odd current is divergence-free in the bulk, $q_\alpha D^{(\mathrm{odd})}_{\alpha\beta} q_\beta = 0$, where $\mathbf D^{(\mathrm{odd})}$ denotes the antisymmetric part of the diffusion matrix. As a consequence, odd diffusivity does not contribute to the decay rate of bulk density modes, since $\partial_t n(\mathbf q,t) = - i \mathbf q \cdot \mathbf J(\mathbf q,t) = - i \mathbf q \cdot \mathbf J^{(\mathrm{even})}(\mathbf q,t)$. Its primary signature is therefore not a modification of bulk relaxation, but rather the emergence of circulating currents near boundaries, which impose constraints on $\mathbf J$.

We remark that viscous contributions introduce additional terms in the diffusion matrix (Eq.~\eqref{dxx}), as derived in Appendix~\ref{app:ficklaw_oddviscosity}. These terms originate from momentum currents generated by the interplay between kinematic and odd kinetic viscosity and the torque density, and they contribute to the density dynamics. However, they appear as higher-order, short-wavelength corrections proportional to $q^4$ and can therefore be safely neglected in the long-wavelength limit.

\section{Linear stability analysis of the underdamped equations of motion}\label{sec:linearStability}

In this section, we demonstrate that the hydrodynamic theory (Eqs.~\eqref{eq:hydro_all}) is linearly unstable around a homogeneous solution when odd viscosity and the torque density strength induced by chirality are large. This analysis suggests that our theory is consistent with the BIO phase (bubble induced by odd interactions) numerically observed \cite{caprini2025bubble}.

To proceed analytically, we consider hydrodynamics in the Fourier space, by applying the Fourier transform to the density $\hat{n}(\mathbf{q}, t)$ and velocity fields $\hat{\mathbf{u}}(\mathbf{q},t)$, where $\mathbf{q}$ is a Fourier space vector (see Appendix~\ref{app:FourierAppendix} for the definitions).
We decompose $\hat{\mathbf{u}}(\mathbf{q},t)$ along the longitudinal $\hat{\mathbf e}_L={\mathbf q}/{q}$ and transverse components $
\hat{\mathbf e}_T=\hat{\mathbf z}\times \hat{\mathbf e}_L$ of the wave vector $\mathbf{q}$. This gives $\hat{u}_L(\bq,t)=\hat{\bu}(\bq,t)\cdot \hat{\mathbf e}_L$ and $\hat{u}_T(\bq,t)=\hat{\bu}(\bq,t)\cdot \hat{\mathbf e}_T$. 
By linearizing our hydrodynamic theory around a homogeneous state with constant density $n_b$ and vanishing velocity, our hydrodynamic theory can be expressed in a compact form as 
\begin{equation}
\hspace{-2.5cm} 
\partial_t
\left( \begin{array}{c}
\delta\hat{n}(\mathbf{q},t)\\ 
\hat{u}_L(\bq,t)   \\ 
 \hat{u}_T(\bq,t) \\ 
\end{array} \right)
= 
\left( \begin{array}{ccc}
0  & -i n_b q & 0\\ 
-i q ( \frac{c_s^2}{n_b}+\frac{\kappa}{m} q^2)&-\gamma- (\nu + \nu_b) q^2& -\nu_{o} q^2   \\ 
 -i q\frac{\chi_o}{m}  &\nu_{o} q^2 &-\gamma-  \nu q^2  \\ 
\end{array} \right) \cdot
\left( \begin{array}{c}
\delta\hat{n}(\mathbf{q},t)\\ 
\hat{u}_L(\bq,t)   \\ 
 \hat{u}_T(\bq,t) \\ 
\end{array} \right) \,\rm{,}
\label{raph1}
\end{equation}
where  $\delta \hat{n}= \hat{n} - n_b$ corresponds to the density fluctuations around the homogeneous state.
Compared to a conventional (passive) fluid, longitudinal and transverse components of the velocity field $\hat {u}_L$ and $\hat{u}_T$ are coupled through odd viscosity, and additionally $\hat{u}_T$ is affected by the density gradient through the strength of the torque density $\chi_o$.
This coupling is naturally associated with the transverse flux generated by the torque density, thereby affecting $\mathbf{u}_T$ and not $\mathbf{u}_L$. 

\begin{figure}
\includegraphics[width=\linewidth,clip=true]{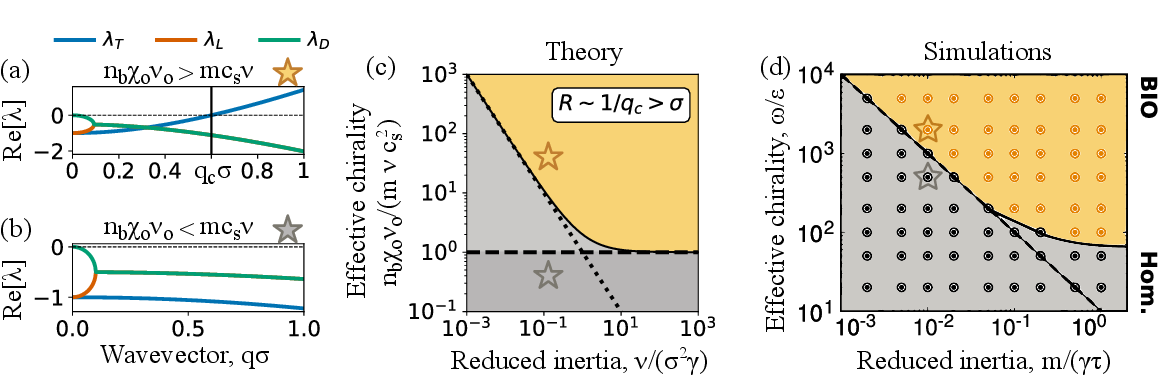}
\caption{\textbf{Linear instability induced by the interplay between odd viscosity and torque density strength}. 
(a)-(b) Real parts of the eigenvalues $\text{Re}[\lambda_T]$ (blue), $\text{Re}[\lambda_L]$ (red), and $\text{Re}[\lambda_D]$ (green) as  functions of the wavevector magnitude $q$, normalized by the particle diameter $\sigma$. 
Panels (a) and (b) correspond to parameter regimes in which the homogeneous state is linearly unstable and stable, respectively.
Specifically, the effective chirality is set to $n_b \chi_o \nu_o/(m \nu c_s^2) = 40$ in panel (a) and $0.4$ in panel (b), while the reduced inertia is fixed to $\nu/(\sigma^2 \gamma) = 0.13$ in both cases.
The horizontal dashed line serves as a guide to the eye indicating the zero value, while the vertical line in panel (a) marks the critical wavevector $q_c$ at which the instability sets in, i.e., where $\text{Re}[\lambda_T]$ becomes positive.
(c) Phase diagram in the plane defined by the reduced inertia, $\nu/(\sigma^2 \gamma)$, and the effective chirality, $n_b \chi_o \nu_o /(m \nu c_s^2)$. The reduced inertia is given by the ratio of the inertial time $1/\gamma$ to the viscous time $\sigma^2/\nu$. The effective chirality corresponds to the product of the torque-density strength $\chi_o$ and the odd viscosity $\nu_o$, appropriately normalized.
Colors in the phase diagram indicate homogeneous phases with vanishing velocity: stable (grey) and unstable (yellow). The two colored stars mark the parameter values used in panels (a) and (b). Dashed black lines in panel (c) serve as guides to the eye: the horizontal line corresponds to $1 = n_b \chi_o \nu_o/(m \nu c_s^2)$, which is the necessary condition for the linear instability, while the diagonal line denotes the condition $q_c \sigma = 1$ (Eq.~\eqref{eq:stability_condition_2}), ensuring that the instability occurs on a length scale larger than the particle diameter $\sigma$.
(d) Phase diagram in the plane of reduced inertia ($\propto 1/\gamma$) and chirality, i.e. the strength, of odd interactions ($\propto \chi_o$), obtained from particle-based simulations of Eqs.~\eqref{Langevin1}.
Here, the color code is the same as in panel (c): grey denotes the homogeneous phase, while yellow indicates the inhomogeneous phase, termed BIO (bubbles induced by odd interactions). The remaining dimensionless parameters used in panels (a)–(c) are $\nu_b/\nu = 0$, $\kappa n_b^2/(m c_s^2) = 0$, $\nu_o/\nu = 1$, and $n_b \nu^2/c_s^2 = 1$.
Panel (d) is adapted with permission from Ref.~\onlinecite{caprini2025bubble}; copyright (2025) AIP Publishing. }
\label{Fig:stability}
\end{figure}

To study the stability of the homogeneous state, we calculate the three eigenvalues associated with the dynamical matrix appearing in Eq.~\eqref{raph1}. Specifically, we look for a solution of the form 
\begin{equation}
\hspace{-2.5cm} 
\left( \begin{array}{c}
\delta\hat{n}(\mathbf{q},t)\\ 
\hat{u}_L(\bq,t)   \\ 
 \hat{u}_T(\bq,t) \\ 
\end{array} \right)
= \left( \begin{array}{c}
\delta\hat{n}(\mathbf{q},0)\\ 
\hat{u}_L(\bq,0)   \\ 
 \hat{u}_T(\bq,0) \\ 
\end{array} \right)
e^{\lambda(\mathbf{q}) t} \,\rm{,}
\label{eq:sol}
\end{equation}
where $\lambda(\mathbf q)$ is one of the three eigenvalues of the dynamical matrix. The homogeneous state is stable when all the eigenvalues have negative real parts $\text{Re}[\lambda](\mathbf{q})<0$. By contrast, if at least one eigenvalue possesses a positive real part for a given $\mathbf{q}$ interval, the homogeneous state is linearly unstable. 
As in a conventional fluid, this eigenvalue problem is described by a cubic secular equation, admitting three exact solutions denoted as $\lambda_D$, $\lambda_L$, and $\lambda_T$.

The eigenvalue $\lambda_{D}$ is the diffusive mode, corresponding to the zero mode related to mass conservation and its redistribution. The eigenvalue $\lambda_{L}$ is a stable mode that has longitudinal polarization in the absence of odd viscosity.
Finally, $\lambda_T$ reduces to the mode of transverse moment diffusion (i.e.\, velocity polarized orthogonal to $\bq$) for vanishing torque density ($\chi_o=0$).
In the absence of odd viscosity ($\nu_o=0$), the homogeneous state is always stable: the dispersion relation of the system is normal, and the real part of the eigenvalues is always negative, as one can see from the exact solution of the secular equation in this limit (See Appendix.~\ref{app:DetailsLinearStability}).

Including a non-vanishing $\chi_o$, we find that either $\lambda_D$ or $\lambda_T$ may become positive when the product between torque density strength and odd viscosity, $\chi_o \nu_o$, is sufficiently large (Fig.~\ref{Fig:stability}(a)), while for low values $\chi_o$ and/or $\nu_o$ the negative sign is restored as in a conventional fluid (Fig.~\ref{Fig:stability}(b)).
As shown in the plot, the change of sign for $\lambda_T$ occurs at finite wavevector $\mathbf{q}>0$, implying that this linear instability emerges at a typical length scale $\sim 1/\mathbf{q}$.
This introduces a natural cutoff on $\mathbf{q}$, that is $\mathbf{q}<\pi/\sigma$, where $\sigma$ is the particle diameter (set by the microscopic repulsive potential).
Thus, we consider unstable only those configurations of parameters where the real part of $\lambda_T$ (or another eigenvalue) becomes positive for $\mathbf{q}\lesssim 1/\sigma$.

The results of the hydrodynamic theory are reported as a function of the model parameters in a phase diagram constructed in the plane of inertia and chirality (Fig.~\ref{Fig:stability}(c)).
The former corresponds to the inertial time $1/\gamma$ normalized by the viscous time scale $\sigma^2/\nu$. 
Even if the latter is identified as the strength of torque density $\chi_o$, we report the product between $\chi_o$ and odd viscosity $\nu_o$ in the phase diagram. Indeed, as it will be confirmed analytically, the stability condition involves the product between $\chi_o$ and odd viscosity $\nu_o$.
The increase of chirality at fixed inertia induces an instability of the homogeneous state, if chirality overcomes the threshold $\chi_o \nu_o n_b > m \nu c_s^2$ (region above the dashed black line in Fig.~\ref{Fig:stability}(c)).
In addition, the phase diagram allows us to conclude that inertia favors the observed instability with an almost linear behavior before a saturation occurs. 
These theoretical findings reproduce qualitatively the numerical results of Ref.~\onlinecite{caprini2025bubble}, obtained through particle-based simulations of the dynamics~\eqref{Langevin1} (Fig.~\ref{Fig:stability}(d)). In correspondence with the inhomogeneous phase (yellow region in Fig.~\ref{Fig:stability}(d)), termed BIO in Ref.~\onlinecite{caprini2025bubble}, our theory predicts a linear instability of the homogeneous state (yellow region in Fig.~\ref{Fig:stability}(c)). Even if this analysis is not proof of the emergence of bubbles, our theory supports their existence.

\subsection{Small wavevector expansion}

Although the eigenvalue problem can be solved exactly, we do not report the lengthy analytical expressions, solutions of Cardano's equation. 
However, we discuss two interesting limits that shed light on the instability numerically observed.
At first, we report the expression for the real part of the three eigenvalues, expanded in powers of $q^2$:
\begin{subequations}
\bea
&&
\lambda_{D}\approx-\frac{c_s^2}{\gamma} q^2  +\mathcal O(q^4),
\\
&&
\lambda_{L}\approx-\gamma
-\left(\nu+\nu_b -\frac{c_s^{2}}{\gamma}\right)q^2-
\frac{\nu_o\bigl(m\gamma\nu_o+n_b\chi_o\bigr)}{m|c_s^{2} - \gamma \nu_b|}
q^2+\mathcal O(\epsilon^3 q^2)+\mathcal O(q^4),\\
&&\label{trasversoodd}
\lambda_{T}\approx -\gamma
-\nu q^2+ \frac{\nu_o\bigl(m\gamma\nu_o+n_b\chi_o\bigr)}{m|c_s^{2} - \gamma \nu_b|}q^{2} +\mathcal O(\epsilon^3 q^2)+\mathcal O(q^4)\,\rm{,}
\label{eigenvaluepassivesmall}
\eea
\label{eigenvalue_expanded}
\end{subequations}
where we have performed an additional expansion in small chirality compared to viscous and pressure terms with $\epsilon={4\nu_o\left(m\gamma^2\nu_o+\chi_o\gamma n_b\right)}/{m(c_s^2-\gamma\nu_b)^2}\ll 1$ and neglecting orders $\epsilon^3$. Details are given in Appendix~\ref{app:DetailsLinearStability}.
In the limit of vanishing $\mathbf{q}$, the diffusive eigenvalue $\lambda_D$ has a vanishing real part, while $\lambda_L$ and $\lambda_T$ assume the value $-\gamma$, consistent with a conventional fluid. 

At this order, chirality affects
$\mathrm{Re}[\lambda_L]$ and $\mathrm{Re}[\lambda_T]$, through an additional $q^2$ contribution. For $\chi_o\nu_o>0$, this chiral correction decreases the value of $\mathrm{Re}[\lambda_L]$, thereby further stabilizing the longitudinal mode, while it increases the value of $\mathrm{Re}[\lambda_T]$, indicating that the transverse mode might become unstable.
Owing to the perturbative character of this expansion, which is valid only at small $q$ and $\nu_o$, we cannot determine the stability of the system at the finite wavevectors where the instability actually occurs. Nevertheless, this trend provides a first theoretical hint of the numerically observed instability at large $\mathbf q$.

\subsection{Large wavevector expansion: finding the instability conditions for the homogeneous state}

To prove the existence of an instability, we consider a large $\mathbf{q}$ expansion, which is performed in the case $\kappa=0$, for simplicity (See Appendix.~\ref{app:DetailsLinearStability} for a discussion on this approximation).
If $\text{Re}[\lambda](q\to\infty)>0$, we can identify a range of parameters where the instability could take place.
The resulting expressions for the three eigenvalues are
    \begin{subequations}
    \bea
    &&\lambda_1(q\to\infty)=\dfrac{-2\nu - \nu_b+\sqrt{\nu_b^2-4\nu_o^2}}{2} q^2,\label{eq:lamnda1}\\ 
    &&\lambda_2(q\to \infty)=\dfrac{-2\nu - \nu_b- \sqrt{\nu_b^2-4\nu_o^2}}{2}q^2,\label{eq:lamnda2}\\ 
    &&\lambda_3(q\to\infty)=\dfrac{\chi_o\nu_on_b-mc_s^2\nu}{m(\nu_o^2+\nu(\nu + \nu_b))}+\mathcal{O}(q^{-2})\label{eq:lamnda3} \,\rm{.}
    \eea
    \end{subequations}
where we label these asymptotic branches as $\lambda_{1}$, $\lambda_{2}$, and $\lambda_{3}$ because, depending on the parameters, they cannot be unambiguously identified with $\lambda_{D}$, $\lambda_{L}$, or $\lambda_{T}$ over the full parameter space since mode crossings and re-ordering may occur. Details are given in Appendix.~\ref{app:DetailsLinearStability}.
Both $\text{Re}[\lambda_1(q\to \infty)]$ and $\text{Re}[\lambda_2(q\to \infty)]$ are negative and decrease as $\mathbf{q}^2$, consistently with our numerical findings. By contrast, $\lambda_3$ -- which can be identified with $\lambda_T$ in Fig.~\ref{Fig:stability}, for the parameter set considered -- saturates to a constant value, which can change sign when the product $\chi_o\nu_o$ is sufficiently large compared to conventional viscosity and equilibrium sound speed. This criterion provides a sufficient condition to observe a linear instability:
\begin{equation}
\label{eq:stability_condition1}
    \chi_o\nu_o n_b> m c_s^2 \nu \,.
\end{equation}
The condition~\eqref{eq:stability_condition1} is consistent with our theoretical phase diagram and is marked by the dashed, horizontal line in Fig.~\ref{Fig:stability}(c). As mentioned earlier, the instability makes sense when it occurs at a length scale larger than the particle diameter, $\sigma$, i.e.\ for $\mathbf{q}$ sufficiently small. 
This requirement stabilizes a region of the theoretical phase diagram (Fig.~\ref{Fig:stability}(c)) even when the condition~\eqref{eq:stability_condition1} is satisfied.

To predict this behavior, we assume again $\kappa=0$ and denote as $\mathbf{q}_c$ the wavevector at which $\lambda_3(q_c)=0$.
By calling $\mathbf{M}$ the dynamical matrix given by Eq.~\eqref{raph1}, the secular equation for the eigenvalues $\det(\mathbf{M}(q_c)-I\lambda(q_c))$ reduces to $\det(\mathbf M(q_c))=0$.
This is an equation determining $q_c$, with solution:
\begin{equation}
q_c=\sqrt{\dfrac{mc_s^2\gamma}{ \chi_o\nu_on_b-mc_s^2\nu}} \,\rm{.} 
\label{eq: qc main text}
\end{equation}
By imposing $\sigma q_c<1$, we obtain our second stability condition
\begin{equation}
    \chi_o\nu_on_b>mc_s^2(\nu+ \gamma\sigma^2)\,\rm{,}
    \label{eq:stability_condition_2}
\end{equation}
accounting both for the oblique line ($\chi_o\nu_on_b>mc_s^2 \gamma\sigma^2$) and the horizontal line ($\chi_o\nu_on_b>mc_s^2\nu$).

Stokes friction, conventional viscosity, and pressure suppress velocity and density gradients; it is therefore not surprising that increasing $\gamma$, $\nu$, and $c_s^2$ favors the uniform state. Here, we focus instead on the destabilizing role of the chiral terms, which requires both the torque density strength $\chi_o$ and the odd viscosity $\nu_o$ to be nonzero and to have the same sign.
To elucidate the underlying mechanism, we describe a physical process leading to the instability of the homogeneous state.
(i) When a density gradient develops, it first generates a transverse velocity perturbation $u_T$ through the torque density term $\chi_o \hat{\mathbf{z}} \times \bfnabla / m$. If $\nu_o = 0$, this perturbation remains purely transverse and is simply damped by viscosity and friction.
(ii) When $\nu_o \neq 0$, odd viscosity couples transverse and longitudinal velocity components, converting the induced $u_T$ into a longitudinal perturbation $u_L$.
(iii) Through the continuity equation, $u_L$ produces additional density gradients, which in turn generate transverse momentum $u_T$ via the torque density $\propto \chi_o$, thereby closing a feedback loop.
An instability arises when this feedback overcomes pressure and dissipative relaxation. If either $\chi_o$ or $\nu_o$ vanishes, the loop is broken and no growth occurs; if $\chi_o$ and $\nu_o$ have opposite signs, the feedback becomes restorative rather than amplifying.

Overall, our theory supports the numerical findings of Ref.~\onlinecite{caprini2025bubble} and the discovered BIO phase. The macroscopic theory developed here identifies torque density and odd viscosity as the two necessary mechanisms to induce the linear instability of the uniform state with constant density and vanishing momentum.

\section{Self-consistent solution of the non-linear equation for a single cavity in the inviscid limit}
\label{sec:nonlinear}

A chiral active fluid subject to odd (transverse) interactions undergoes a transition from a uniform state to a non-homogeneous phase characterized by bubbles, namely regions devoid of particles that spontaneously form even in the absence of attractive interactions. This novel phase, termed BIO (bubbles induced by odd interactions), arises as a direct consequence of chirality-induced transverse forces. In general, bubbles may nucleate due to density fluctuations generated by thermal noise. However, whereas in a standard fluid, void regions are unstable and are continuously created and destroyed, the presence of odd interactions stabilizes these voids, allowing them to reach a finite size.

In the previous section, we showed that the hydrodynamic equations governing a chiral active fluid become linearly unstable when chirality and odd viscosity are sufficiently large. Although this linear stability analysis supports the numerical findings of Ref.~\onlinecite{caprini2025bubble}, it cannot be regarded as a direct proof of the BIO phase, since it does not demonstrate that the chirality-induced inhomogeneous state is specifically characterized by the formation of bubbles.

In Ref.~\onlinecite{caprini2025bubble}, the mechanism leading to bubble formation was discussed intuitively using a scaling argument. When a small bubble (i.e., a density inhomogeneity) forms, the odd interactions become unbalanced, generating net tangential forces -- absent in equilibrium -- that drive particles along the bubble interface (edge currents). Additionally, each particle performing circular motion experiences a centrifugal force acting normal to the bubble surface, which increases with the strength of odd interactions. Bubbles become stable when this centrifugal force is sufficiently strong to balance the pressure arising from the repulsive interactions. The interplay between these two effects gives rise to a stable, non-uniform macroscopic phase characterized by nearly circular voids embedded in a liquid.

This argument provides a qualitative explanation for bubble stability in the BIO phase, but it does not constitute a macroscopic theory. To address this gap, we now present a special solution of the steady hydrodynamic equations governing a chiral active fluid (Eq.~\eqref{eq:hydro_all}). This solution describes a vortex encircling an empty region, closely resembling the BIO phase observed in numerical simulations.

Here, we start with the stationary Navier–Stokes equation developed for a chiral active fluid~\eqref{velocita}, where we have neglected the time derivative of the velocity field:
  \bea
  (\bu\cdot \bfnabla)\bu(\rr)=  \frac{1}{m n}\bfnabla \cdot {\boldsymbol \sigma}^{(sym)} (\rr)- \gamma \bu(\rr)-\frac{\chi_o} {m} (\hat{\bf z} \times \bfnabla n(\rr)) \,\rm{.}
\label{lasteq}
\end{eqnarray} 
To proceed, for simplicity, we neglect viscous effects by setting $\nu_b = 0$, $\nu = 0$, and $\nu_o = 0$ which yields: ${\boldsymbol \sigma}^{(sym)} =-\bfnabla P(\rr)  +\kappa n(\rr) \bfnabla \bfnabla^2 n(\rr))$. 
We remark that viscous terms play a crucial role in inducing the instability of the homogeneous state. However, here, our aim is not to describe the onset of the linear instability (Sec.~\ref{sec:linearStability}), but to address the final steady state. We thus consider a fully non-linear regime, where bubbles arise from the balance between the pressure and the inertial contributions generated by the advective term. 

Particle-based simulations of the homogeneous and BIO phases reported in Ref.~\onlinecite{caprini2025bubble} suggest that we should seek steady bubble-like solutions with an almost circular shape and circulating currents on the surface of bubbles.
For simplicity, we restrict to a single bubble and introduce polar coordinates, the radial position $r$ calculated from the bubble center $r=0$, and the azimuthal angle $\theta$. In these coordinates, the velocity field has an axisymmetric, purely azimuthal profile:
\begin{equation}
    \bu = u_\theta(r) \bm e_\theta \,\rm{,}
    \label{eq:AnsatzRadial}
\end{equation}
where $\mathbf{e}_\theta$ is the unit vector along the tangential direction.
In other words, $u_r = 0$ and $u_\theta$ depend only on the radial distance $r$. Although solutions corresponding to multiple cavities are, in principle, possible, they are not considered here.

Inserting the bubble Ansatz Eq.~\eqref{eq:AnsatzRadial} into the odd Navier-Stokes equation, we obtain two ordinary differential equations by projecting on the tangential and radial unit vectors, $\bm e_\theta$ and $\bm e_r$:
\begin{subequations}
\bea&&
m\gamma u_\theta=-\chi_o \partial_r n(r),
 \label{Tangentialbalance}
 \\&&
-\frac{u_\theta^2}{r}=
-\frac{1}{mn} \partial_r P +\dfrac{\kappa}{m} \left( \frac{\partial^3 n }{\partial r^3}  +\frac{1}{r}  \frac{\partial^2 n}{\partial r^2}  
-  \frac{1}{r^2}  \frac{\partial n}{\partial r}  \right) \,\rm{.}
\label{Radialbalance}
 \eea
 \end{subequations}
Equation~\eqref{Tangentialbalance} implies that a non-vanishing tangential velocity is induced by a radial density gradient at the bubble interface and is balanced by damping. As expected, when $\partial_r n$ vanishes, as in a homogeneous state, one has $u_\theta=0$. Finally, by construction, this velocity field is divergence-free, $\bfnabla\cdot\bu=0$, making the flow nearly incompressible and ensuring that the stationary continuity equation is satisfied because $\bu$ and $\bfnabla n$ are perpendicular. 
Equation~\eqref{Radialbalance} expresses the radial force balance: the outward centrifugal term $m u_\theta^{2}/r$, which tends to expand the bubble, is counteracted by the inward pressure gradient and the surface tension, controlled by~$\kappa$. For simplicity, we neglect viscous contributions, which primarily act to smooth the velocity profile (see Appendix~\ref{polarcoordinates}).
Substituting Eq.~\eqref{Tangentialbalance} into Eq.~\eqref{Radialbalance} yields:
 \beq
\frac{\chi_o^2}{m\gamma^2}\,\frac{n(r)}{r} (\partial_r n(r))^2
=\frac{\partial P}{\partial r}-\kappa n(r) \frac{\partial}{\partial r} \left( \nabla^2 n(r)\right)\,\rm{.}
\label{equazioneradialea}
\eeq
Equation~\eqref{equazioneradialea} describes the density profile around an arbitrary point in the system (see Appendix~\ref{polarcoordinates}).
To proceed, we assume that the local pressure depends on the spatial coordinates through the density and an equation of state, accounting for volume exclusions through short-range repulsive interactions. 
In other words, the bulk pressure is a monotonically increasing function of the density in the entire range of density. We note that although the fluid appears quasi-incompressible, this property emerges from the chosen velocity profile and its coupling to chirality, i.e., the strength of torque density. Consequently, the pressure remains a genuine dynamical field, rather than a constraint that must be tuned to enforce incompressibility.
We approximate the derivative of the pressure by:
$
\frac{\partial P}{\partial r}\approx (\frac{d P}{d n})\, \partial_r n(r)
$
and obtain the following equation 
\beq
\kappa \left(  \frac{\partial^3 n }{\partial r^3}+\frac{1}{r}  \frac{\partial^2 n}{\partial r^2}  \right)+
\frac{\chi_o^2}{m\gamma^2}\,\frac{1}{r} \left(\frac{\partial n}{\partial r}\right)^2
- \frac{1}{n}\left(\frac{d P}{d n}\right)\, \left(\frac{\partial n}{\partial r}\right)-\kappa  
  \frac{1}{r^2}  \frac{\partial n}{\partial r}=0   \,\rm{.}
\label{fbbeqf37a}
\eeq
This equation always possesses solutions with $\bfnabla n={\bf0}$ describing a fluid with $\bu={\bf0}$ and constant density, i.e., a uniform phase, but also a vacuum with $(n, \bu)=(0, \bf{0})$. Of course, the particle number conservation
$\int d\rr n(\rr)=N$
forbids an empty system, but it does not forbid the formation of a non-uniform phase characterized by the presence of an empty and a dense region, i.e., a bulk fluid with a void. 

\begin{figure}
\includegraphics[width=0.8\linewidth,clip=true]{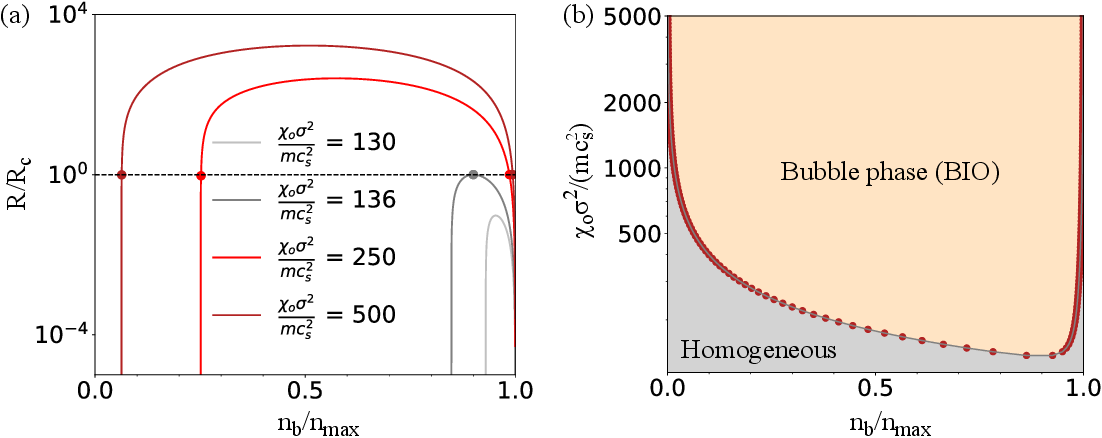}
\caption{{\bf Non-linear phase diagram}
(a) Predicted bubble radius $R_c$ (Eq.~\eqref{eq:Rcn_expr}) as a function of the bulk average number density $n_b$ for different values of the chirality, quantified by the torque-density strength $\chi_o$. The horizontal dashed line indicates the threshold used for bubble identification, set to $R_c = \sigma$, where $\sigma = 1$ is the particle diameter. (b) Phase diagram in the plane of number density $n_b$ and chirality, represented by the torque-density strength $\chi_o$. Colors denote the different phases: the homogeneous phase (grey) and the BIO phase~\cite{caprini2025bubble} (yellow), corresponding to bubbles induced by odd interactions. Red points indicate configurations for which $R_c = \sigma$, corresponding to the minimal radius used to identify a bubble. Consequently, the phase diagram is constructed using the straight-line construction illustrated in panel (a). The panels are obtained from Eq.~\eqref{eq:Rcn_expr} using the parameters $\sigma \gamma / c_s = 100$, $\ell/\sigma = 0.1$, $T /(m c_s^2) = 0.02$, and $\kappa /(m \sigma^4 c_s^2) = 10$, where $m = 1$ is the particle mass and $c_s = 1$ is the sound speed of the fluid. We note that all kinematic viscosities are set to zero, $\nu_b = \nu = \nu_o = 0$.
}
\label{Fig:instability}
\end{figure}

Hereafter, we use the hydrodynamic equation~\eqref{fbbeqf37a} to build a solution with a non-trivial density profile ($\bfnabla n(\rr)\neq {\bf0}$) which corresponds to a single circular cavity centered at the origin of the coordinate system and surrounded by a fluid and an edge current tangential to the cavity surface ($u_\theta\neq 0$), in other words a vortex.

We can simplify Eq.~\eqref{fbbeqf37a} for a system characterized by an empty region (bubble), where a circular domain of radius $R_{c}$ and density $n_{e} \approx 0$ is surrounded by a bulk with density $n_{b}$. We describe the system by placing the origin of a polar coordinate system at the center of the circular empty region, i.e., the bubble.
The radial density profile therefore increases asymptotically from $n_{e}$ ($r \ll R_{c}$) to $n_{b}$ ($r \gg R_{c}$), reaching at the interface an intermediate value $n_{\text{int}} = (n_b + n_e)/2$ over a width $\ell$, defined by $\ell^2 = \kappa n / ({dP}/{dn})|_{n=n_{\text{int}}}$. We can suppose that $\ell$ does not depend on the strength of the chirality-induced torque density $\chi_{o}$, but solely on the bulk and bubble densities: the introduction of this auxiliary length accounts for a smoothing of the radial density profile.
Using the interface density scale, we can reduce the order of the differential equation~\eqref{fbbeqf37a} by performing the following substitutions: $\partial_r n = \phi {n_b}/{\ell}$ and $r \to \ell z$. Since density gradient are strong only close to the interface, we approximate the pressure term by its value at the interface, $({dP(r)}/{dn})/n(r) \approx ({dP(r)}/{dn})/n(r)|_{n=n_{\text{int}}}$. Therefore, Eq.~\eqref{fbbeqf37a} can be rewritten as
\bea
\partial^2_z \phi +\frac{1}{z}\partial_z \phi
-\frac{1}{z^2} \phi - \phi  =-\frac{\chi_o^2}{ m\gamma^2}\, \frac{n_b} { \kappa }  \,\frac{1}{z} (\phi(z))^2 \,\rm{.}
\label{eq:homovortexa}
\eea
The left-hand side of Eq.~\eqref{eq:homovortexa} has the form of the modified Bessel differential equation of integer order $1$ whose physical solution decays exponentially at infinity~\footnote{ The Bessel differential equation is
$$
 \frac{d^2 B_k}{dz^2} +\frac{1}{z} \frac{d B_k}{dz} -\left(1+\frac{k^2}{z^2}\right)B_k=0
 $$
 and in our case the index $k=1$.}
 $
 \phi(z)\propto K_1(z)\approx\sqrt{\frac{\pi}{2z}} e^{-z}(1+\mathcal{O}(|z|^{-1})\,)
 $.
Even if the nonlinearity in Eq.~\eqref{eq:homovortexa} precludes an exact solution, we remark that we can still extract useful information about the thickness of the interface, $\ell$.
Indeed, the function $\phi(z)$ reaches zero exponentially far away from the bubble surface $r\gg R_c$. In this limit, the non-linear term on the right-hand side of Eq.~\eqref{eq:homovortexa} can be neglected, and $\phi(z)$ can be approximated as 
 \begin{equation}
    \phi(z)=\frac{K_1(r/\ell)}{K_1(R_{c}/\ell)}\approx  \sqrt{\frac{R_c}{r}} e^{-(r-R_c)/\ell} \,\rm{,}  
    \label{Eq:Asympotitc_Sol}
 \end{equation}
 for $r\gg R_c$.
This result holds only far from the interface. For $r \lesssim R_c$, integrating $n=\mathrm{const}+n_b\int (\phi(r)/\ell)dr$ would make $n(r)$ decrease and eventually become negative. This unphysical behavior would be cured by the exact solution of Eq.~\eqref{eq:homovortexa}, which is analytically intractable. Instead, guided by the asymptotic form in Eq.~\eqref{Eq:Asympotitc_Sol}, we propose the following global ansatz for the density profile:
\begin{equation}
    n(r)= \frac{n_b}{1+\exp\left(-\dfrac{r-R_c}{\ell}\right)}\,\rm{.}
\label{eq:radial_density_profile}
\end{equation}
This sigmoid form, widely used in cavitation and nucleation theory~\cite{onuki2002phase}, smoothly interpolates between $n(r\ll R_c)=0$ and $n(r\gg R_c)=n_b$ with an interface width proportional to $\ell$. 
By integrating Eq.~\eqref{equazioneradialea}:
\bea
P(n(r\to\infty)=n_b)-P(n(r\to0)=0)= \dfrac{\chi_{o}^{2}}{m\gamma^{2}}\int_{0}^{\infty}\dfrac{dr}{r}\left(\partial_{r}n(r)\right)^{2}\left(n(r)  - \dfrac{m\kappa\gamma^{2}}{\chi_{o}^{2}}\right)\,\rm{,}
\label{eq:IntegroDiff_Pn_relation}
\eea
we can find $R_c$ self-consistently 
by substituting the radial density profile Eq.~\eqref{eq:radial_density_profile} into Eq.~\eqref{eq:IntegroDiff_Pn_relation}, obtaining
\bea
R_c = \frac{\chi_o^2n_b^2}{12m\gamma^2\ell}
\left(n_b - 2\dfrac{m\gamma^2}{{\chi_o^2}}\kappa\right)\frac{1}{P(n_b)}\,\rm{.}
\label{eq:Rcn_expr}
\eea
The bubble radius $R_c$ increases monotonically with chirality, i.e., with the strength of the torque density $\chi_o$, and decreases as the friction coefficient $\gamma$ grows. In addition, $R_c$ exhibits a non-monotonic dependence on the average particle density $n_b$, independent of $\chi_o$ (Fig.~\ref{Fig:instability}(a)). 
This non-monotonicity is rooted in the dependence of the pressure $P(n_b)$ with $n_b$. Indeed, independently of the closure for $P(n_b)$, the pressure linearly increases with $n_b$ in the dilute limit (ideal gas pressure) while it diverges for large $n_b$ due to volume-exclusion effects. 
This is the case of Henderson’s closure for the pressure~\cite{henderson1975simple}:
\bea
P(n)=nk_{B}T
\dfrac{1+\dfrac{1}{8}\left(\dfrac{\pi}{4}\sigma^{2}n\right)^{2}}
{\left(1-\dfrac{\pi}{4}\sigma^{2}n\right)^{2}}\,\rm{,}
\label{eq:Hendersons_pressure}
\eea
which we adopt as a representative case study.
In this specific example, the critical radius $R_c$ grows quadratically with $n_b$ in the dilute limit at low $\kappa$, while it decreases for large $n_b$.
We do not expect Henderson’s equation of state to be quantitatively accurate in the dense regime. In particular, expression~\eqref{eq:Hendersons_pressure} predicts a divergence at $n\sigma^2\pi/4 = 1$, whereas hard disks approach ideal (or random) close packing at significantly lower densities and may undergo crystallization, which is not accounted for by this closure.
For soft disks, one likewise expects the pressure to increase very steeply rather than to diverge at a sharply defined density. Nevertheless, independently of the specific equation of state, there must exist a characteristic density $n_{\max}$ beyond which the pressure rises rapidly as the typical interparticle spacing becomes comparable to $\sigma$. For our illustrative purposes, Eq.~\eqref{eq:Hendersons_pressure} therefore provides a convenient closure that captures this qualitative behavior.

The BIO phase can be identified from Eq.~\eqref{eq:Rcn_expr} as a configuration where the bubble radius exceeds the particle diameter, $R_c>\sigma$. This prediction can be used to construct a phase diagram distinguishing homogeneous and BIO phases. Notably, the analytical expression~\eqref{eq:Rcn_expr} provides results consistent with the numerical findings of Ref.~\onlinecite{caprini2025bubble}, since $R_c\sim \chi_o^2$ and $R_c\sim 1/\gamma^2$, the BIO phase is favored when the strength of torque density or inertia is increased.

In addition, Eq.~\eqref{eq:Rcn_expr} allows us to predict the behavior of the BIO phase as a function of density, going beyond the numerical results reported in Ref.~\onlinecite{caprini2025bubble}. Specifically, in Fig.~\ref{Fig:instability}(b), we show a phase diagram in the plane defined by chirality $\chi_o$, and the average particle density $n_b$. Our analysis reveals a U-shaped phase diagram: in the dilute limit, $n_b \to 0$, particles rarely interact, and the BIO phase is suppressed; conversely, in the dense limit, $n_b \to 1$, the BIO phase is disfavored, requiring larger $\chi_o$ to emerge. Indeed, as density increases, higher torque density values are necessary to overcome the pressure arising from volume exclusion effects.

\section{Conclusions}

In this paper, we derive a macroscopic hydrodynamic theory for an inertial chiral active fluid in two spatial dimensions, governed by Langevin dynamics. The particles interact through short-range repulsive forces and odd interactions, which conserve linear momentum but break both angular momentum and kinetic energy conservation. These additional transverse forces originate from the coarse-graining of rotational degrees of freedom and frictional interactions, allowing the dynamics to be described solely in terms of particle positions and velocities.

Using mean-field approximations, we derive a Boltzmann-like equation for the single-particle distribution of the chiral fluid and, through a velocity-moment expansion, obtain the continuity equation for the particle density and a Navier–Stokes-like equation for the velocity field. The latter relates momentum variations to body forces, friction, and the stress tensor, which is expressed in terms of the density and velocity fields.
The static pressure contribution is approximated using an equation of state consistent with the model’s repulsive interactions, while the viscous stress tensor is described by phenomenological constitutive relations linking stress to velocity gradients through viscous transport coefficients. The breaking of parity symmetry generates odd viscous terms that are symmetric and linear in the velocity gradients, as well as an additional antisymmetric contribution that depends solely on the density gradient and emerges from a gradient expansion. This latter term has the form of a torque density and must be included in a hydrodynamic treatment of chiral fluids.

Within our hydrodynamic framework, we conclude that chirality induces odd diffusion and drives the emergence of an inhomogeneous phase, termed BIO (bubbles induced by odd interactions), in agreement with particle-based simulation results. A linear stability analysis of the hydrodynamic equations shows that the homogeneous state, characterized by uniform density and vanishing velocity, becomes unstable when the conventional (even) viscosity is sufficiently small compared to the odd viscosity and the strength of the chirality-induced torque density.
Beyond the linear stability analysis, our theory admits a steady-state solution describing a circular cavity surrounded by a vortex. This solution can be regarded as an elementary building block of the BIO phase observed in simulations, which consists of multiple voids encircled by circulating currents.

Our theory identifies odd viscosity and torque density as the key macroscopic ingredients required to reproduce the BIO phase. As a consequence, this collective behavior transcends the specific particle model considered here and emerges as a general phenomenon in active matter, distinct from previously known phases.
While alignment or self-alignment interactions give rise to flocking or swarming behavior~\cite{vicsek2012collective,cavagna2014bird,ihle2011kinetic,giavazzi2018flocking,das2024flocking,musacchio2025self}, and persistent self-propelled motion leads to motility-induced phase separation~\cite{cates2015motility,fily2012athermal,buttinoni2013dynamical,solon2015pressure,digregorio2018full,caprini2020spontaneous}, we show here that chirality and the transfer of angular momentum instead generate an inhomogeneous phase characterized by bubbles.

\appendix

\section{Heuristic derivation of the transverse force from a collisional model}
\label{heuristicappendix}

In the present Appendix, we connect our dynamics, characterized by odd interactions, with a model proposed by Digregorio et al.~\cite{digregorio2025phase} to describe granular active spinners.
We consider the encounter of two identical rough rotating disks, labeled $i$ and $j$, with a common diameter $\sigma$.
As they come into contact, their relative velocities are reduced due to friction between their surfaces.
The relative velocity of the surfaces of the colliding
particles at the point of contact is $\bm{v}_{ij}=\bm{v}_j-\bm{v}_i$, while $\bm{v}_{ij}^{\rm{t}}$ denotes its component transverse to the line joining the centers of the particles. The transverse velocity, $\bm{v}_{ij}^{\rm{t}}$, can be expressed in terms of the positions $\bm{r}_{i}$ and $\bm{r}_{j}$, the translational velocities $\bm{v}_{i}$ and $\bm{v}_{j}$ of the disk centers, and the angular velocities $\bm{\omega}_i$ and $\bm{\omega}_j$ about their axes as
\beq
\bm{v}_{ij}^{\rm{t}} = \bm{v}_{ij} - (\bm{v}_{ij} \cdot \hat{\bm{n}}_{ij})\hat{\bm{n}}_{ij} - \bm{\omega}_{ij} \times \hat{\bm{n}}_{ij} \,\rm{,}
\label{transversevelocityrel}
\eeq
where we have defined
$
\bm{\omega}_{ij}=(\bm{\omega}_i+\bm{\omega}_j)\sigma/2
$ and
$
\hat{\bm{n}}_{ij}=(\bm{r}_i-\bm{r}_j)/r_{ij}\,.
$
The velocity-dependent transverse force has a finite range $\sigma$ (a condition enforced by the Heaviside function $\Theta$):
\beq
 \bm{F}_i^{\rm{t}}= -\gamma_{\perp}\Theta\left( \sigma-\lVert\bm{r}_{i}-\bm{r}_{j}\rVert\right)  \bm{v}_{ij}^{\rm{t}}\,\rm{.}
 \label{forcetransv}
\eeq
Here, $\gamma_\perp$ represents a friction coefficient that contributes to the change in linear momentum of particle $i$.
We can now write the equations of motion for the translational and rotational degrees of freedom of the $i$-th disk,
including the torque exerted by particle $j$ and that exerted by a thermal bath on the angular degree of freedom:
\begin{subequations}
\bea
&&\label{eq:motion_tr}
m \frac{{\rm{d}} \bm{v}_i}{{\rm{d}}t} = -m\gamma \ \bm{v}_i + \sqrt{2m\gamma T}\, \boldsymbol{\xi}_i + \mathbf{F}_i^\mathrm{rep} + \bm{F}_i^{\rm{t}} \,\rm{,} \\
&&\label{eq:motion_rot}
I \frac{{\rm{d}}\bm{\omega}_i}{{\rm{d}} t} 
 =  -\gamma_{\theta} \ \bm{\omega}_i  + \sqrt{2\gamma_{\theta}D_{\mathrm{rot}}}\, \boldsymbol{\xi}_{\theta,i} + \boldsymbol{\tau}_i + \boldsymbol{\tau}_0 \,\rm{.}
\eea
\end{subequations} 
Equation~\eqref{eq:motion_tr} is an underdamped equation of motion for the velocity and has the same structure as Eq.~\eqref{Langevin1}. Specifically, it describes a particle moving in a medium with drag coefficient $m\gamma$, thermal fluctuations modeled by the Gaussian stochastic term $\boldsymbol{\xi}_i$ with zero mean and unit variance, and interactions with the $j$-th disk through repulsive ($\bm{F}_i^{\rm{rep}}$) and transverse ($\bm{F}_i^{\rm{t}}$) forces.
Equation~\eqref{eq:motion_rot} describes the evolution of the angular momentum of a disk with moment of inertia $I$, subject to a rotational friction torque $\gamma_{\theta}\bm{\omega}_i$ and exchanging angular momentum with the bath via the stochastic term $\boldsymbol{\xi}_{\theta,i}$. In addition, a constant torque $\bm{\tau}_0$, normal to the plane of motion, is applied to the particle, while the $j$-th particle exerts a net interaction torque $\boldsymbol{\tau}_i$.
We seek a simplification of Eq.~\eqref{eq:motion_rot} by considering the steady-state regime with low rotational diffusion. In this regime, the frictional torque can be absorbed into a redefinition of the constant torque $\boldsymbol{\tau}_0$.
Thus, we have $\boldsymbol{\tau}_i \approx - \boldsymbol{\tau}_0$.
On the other hand, the torque $\boldsymbol{\tau}_i$ is produced by the tangential force $\bm{F}_j^{\rm{t}} = - \bm{F}_i^{\rm{t}}$:
\beq
\bm{\tau}_i=-\frac{\sigma}{2} \hat{\bm{n}}_{ij}\times  \bm{F}_j^{\rm{t}} = \frac{\sigma}{2} \hat{\bm{n}}_{ij}\times  \bm{F}_i^{\rm{t}} \,\rm{.}
\eeq
If we further assume that, in Eq.~\eqref{transversevelocityrel}, the relative translational velocities $\bm{v}_{ij}\approx 0$ are negligible compared to the angular velocities, we obtain a simple relation between $\bm{\tau}_i$ and $\bm{\omega}_{ij}$:
\beq
\label{eq:tau_i_relation}
\bm{\tau}_i\approx \gamma_\perp \sigma^2\, \frac{\Theta\left( \sigma-\lVert\bm{r}_{i}- \bm{r}_{j}\rVert\right) }{4} \,\hat{\bm{n}}_{ij}  \times \Bigl( \bm{\omega}_{ij} \times \hat{\bm{n}}_{ij}\Bigr)
= \gamma_\perp \sigma^2\,  \frac{\Theta\left( \sigma-|\bm{r}_{i}- \bm{r}_{j}|\right) }{4} \bm{\omega}_{ij}\,\rm{.}
 \eeq
Finally, considering Eq.~\eqref{eq:tau_i_relation} and Eq.~\eqref{forcetransv} under the previous approximation for $\bm{v}_{ij}^{\rm{t}}$, we obtain: 
\bea 
\gamma_\perp\Theta\left( \sigma-\lVert\bm{r}_{i}-\bm{r}_{j}\rVert\right) \dfrac{\sigma}{2}\bm{\omega}_{ij} \approx 2\dfrac{\bm{\tau}_i}{\sigma} \approx -2\dfrac{\bm{\tau}_0}{\sigma}\,\rm{.}
\eea
Thus, one can express the transverse force in Eq.~\eqref{forcetransv} in terms of the applied torque $\bm{\tau}_0$:
\beq
\bm{F}_i^{\rm{t}}\approx  -2\dfrac{\bm{\tau}_0}{\sigma}\times \hat{\bm{n}}_{ij}\,\rm{,}
\eeq
which represents a force perpendicular to the normal of the plane of motion and to the vector $\hat{\bm{n}}_{ij}$. Therefore, $\bm{F}_i^{\rm{t}}$ is tangential and independent of the velocities $\bm{v}_{i}$ and $\bm{v}_{j}$.
The above construction can be generalized by replacing the Heaviside function with a generic reciprocal function of the inter-particle distance, for example, a function that decays with $r_{ij}$.

The present derivation of the odd force also clarifies why a system subject to transverse forces can be regarded as active.
Indeed, the odd force provides a coarse-grained description of an active system in which activity originates from the self-propelled rotation of the particles, induced by the stochastic angular equation~\eqref{eq:motion_rot}.
Upon eliminating the angular degrees of freedom, the dynamics reduces to an effective force description in terms of ${\bf F}_{ij}^{\perp}$.

\section{The effective mean odd force}
\label{effectiveoddforce}

In the present Appendix, we derive an explicit expression for the effective transverse force in terms of the density field and the interaction potential.
Using the definition in Eq.~\eqref{pseudopotentialforce}, we consider the effective field generated by a spatial distribution of particles $n(\rr)$:
\beq
\label{eq:f_perp}
{\bf f}^{\perp} (\rr)=\int d\rr' \, \fvec^{\perp}(|\rr-\rr'|) n(\rr')=  
\int d\rr' \,\frac{ d U^{\perp}(R)}{\partial R} ({\hat {\bf z}} \times \ebh)\,  n(\rr') \,.
\eeq
Here we have written $\rr^{\prime}=\rr+ R \ebh$, where $R=\lVert\rr-\rr^{\prime}\rVert$ is the distance between $\rr^{\prime}$ and $\rr$ and $\ebh=(\cos(\theta),\sin(\theta))$ the unit vector defining its direction on the two-dimensional $(x,y)$ plane.
Assuming a slow spatial variation of the density field around the point $\rr$, we perform the following Taylor gradient expansion~\cite{huang2025anomalous}:
\beq
\label{eq:n_expansion}
n(\rr')=n(\rr+ R \ebh) = n(\rr)+R\sum_\alpha \frac{\partial n(\rr)}{\partial r_\alpha}\hat e_\alpha+\frac{1}{2} R^2
\sum_{\alpha\beta} \frac{\partial^2 n(\rr)}{\partial r_\alpha \partial r_\beta}\hat e_\alpha \hat e_\beta+\frac{1}{6} R^3
\sum_{\alpha\beta\gamma} \frac{\partial^3 n(\rr)}{\partial r_\alpha \partial r_\beta \partial r_\gamma}\hat e_\alpha\hat e_\beta \hat e_\gamma+\mathcal{O}\left(R^{4}\right)\,.
\eeq
Substituting Eq.~\eqref{eq:n_expansion} into Eq.~\eqref{eq:f_perp} and noting that all even-power terms vanish after angular integration, we obtain
\beq
 {\bf f}^{\perp} (\rr) =   \int_0^{2\pi} d\theta \int_0^\infty R dR\,   \frac{ d U^{\perp}(R)}{\partial R} ({\hat {\bf z}} \times \ebh)     
 \, \left(R\sum_\alpha  \frac{\partial n(\rr)}{\partial r_\alpha}\hat e_\alpha
+\frac{1}{6} R^3
\sum_{\alpha\beta\gamma} \frac{\partial^3 n(\rr)}{\partial r_\alpha \partial r_\beta \partial r_\gamma}\hat e_\alpha \hat e_\beta \hat e_\gamma + \mathcal{O}\left(R^{5}\right)\right) \,,
\eeq
where, for practical reasons, we retain only the first and second non-vanishing terms in the expansion of $n(\rr^{\prime})$.
Using the properties of $\ebh$, we can express the effective transverse interactions in terms of the vector product of $\hat{\bf z}$ with the gradient of $n(\rr)$ and its Laplacian:
\beq
 {\bf f}^{\perp} (\rr) =  \pi\hat {\bf z}\times \bfnabla \Big[  n(\rr)  \int_0^\infty R^2 dR\,   \frac{ d U^{\perp}(R)}{\partial R}
 + \frac{1}{8}   \nabla^2 n(\rr)  \, \int_0^\infty R^4 dR\,   \frac{ d U^{\perp}(R)}{\partial R} +\mathcal{O}\left(R^{7}\right)\Bigr]\,.
\eeq
Considering only the leading term in the gradient expansion, we obtain
\bea&&
\mathbf{f}^{\perp}\approx -\chi_o \hat{\mathbf{z}}\times \nabla n(\rr)\,,
\label{transversedeltab}
\eea
where we have defined the effective constant $\chi_o$ as
\beq
\chi_o= - \pi  \int_0^\infty R^2 dR\,   \frac{ d U^{\perp}(R)}{\partial R}
\label{Aconstant2}\,.
\eeq
This constant is positive if $\frac{ d U^{\perp}(R)}{\partial R} <0$, within our approximations. We recall that $\mathbf{f}^{\perp}$ depends only on odd-order spatial derivatives of $n$. For instance, at leading order, the effective source term depends on the first spatial derivative of the density field and is non-zero whenever the system exhibits density inhomogeneities.
Finally, we consider the $z$-component of the curl of the effective force:
\beq
\label{eq:vorticity_source}
(\bfnabla \times \mathbf{f}^{\perp}(\rr))_z=\partial_x  \mathbf{f}_y^{\perp}-\partial_y  \mathbf{f}_x^{\perp} \,.
\eeq
Using the expression for $\mathbf{f}^{\perp}$, Eq.~\eqref{eq:vorticity_source} becomes
\beq
(\bfnabla \times {\bf f}^{\perp}(\rr))_z=  \left( \pi \int_0^\infty R^2 dR\,   \frac{ d U^{\perp}(R)}{\partial R} \right)\, (\partial^2_x +\partial^2_y) n(\rr) \,.
\eeq
Equivalently, taking into account the constant $\chi_o$ defined in Eq.~\eqref{Aconstant}, we write
\beq
(\bfnabla \times {\bf f}^{\perp}(\rr))_z= -\chi_o\, (\partial^2_x +\partial^2_y) n(\rr)\,\rm{.}
\label{endofthestory}
\eeq
Therefore, Eq.~\eqref{endofthestory} shows that the effective transverse force ${\bf f}^{\perp}$ produces a non-vanishing torque.

\section{The viscous stress tensor and the odd viscosity}

\label{oddviscoustensor}
For completeness, we briefly recall the notion of odd viscosity, a phenomenon that can arise when both time-reversal symmetry and parity are broken. As with the standard (even) viscosity, deriving it from a microscopic model is beyond our scope, and we therefore treat it as a phenomenological parameter. In the literature, there are very few examples in which odd viscosity is derived directly from the underlying force laws. One such example is the paper by Kaufman~\cite{kaufman1960plasma}, which deals with a fully ionized plasma in a magnetic field.

In general, up to linear order in the velocity gradients, one can express the viscous part of the stress tensor, $\tilde \sigma_{\alpha\beta}$, in terms of the rate of strain tensor, defined as
\bea
u_{\alpha\beta}=\frac{1}{2}\left(\frac{\partial u_\alpha}{\partial x_\beta}+\frac{\partial u_\beta}{\partial x_\alpha} \right)\,\rm{.}
\eea
For instance, one can define the viscous stress tensor as a contraction between a fourth rank tensor $\eta_{\alpha\beta\gamma\delta}$, called the viscosity tensor, and the rate of strain:
\bea
\tilde \sigma_{\alpha\beta}=\sum_{\gamma\delta}\,\eta_{\alpha\beta\gamma\delta}\, u_{\gamma\delta}\,\rm{.}
\eea
Since both the stress and the rate of strain tensors are symmetric under the exchange of $\alpha$ with $\beta$ and $\gamma$ with $\delta$, the viscosity tensor $\eta_{\alpha\beta\gamma\delta}$ must preserve these symmetries.
However, the most general $\eta_{\alpha\beta\gamma\delta}$ can contain terms that are symmetric or antisymmetric under the exchange of the index pair $(\alpha,\beta)$ with the index pair $(\gamma,\delta)$.
One can therefore decompose the viscosity tensor $\eta_{\alpha\beta\gamma\delta}$ as the sum of its even and odd components under such an exchange:
\bea
\eta_{\alpha\beta\gamma\delta}= \eta^{\text{even}}_{\alpha\beta\gamma\delta}+ \eta^{\text{odd}}_{\alpha\beta\gamma\delta}\,\rm{.}
\eea
Avron~\cite{avron1998odd} has shown that, when time-reversal symmetry and parity are broken, for a two-dimensional isotropic system, one can write
\begin{subequations}
\bea&&
\eta^{\text{even}}_{\alpha\beta\gamma\delta}
=
\eta (\delta_{\alpha\gamma}\delta_{\beta\delta}+\delta_{\alpha\delta}\delta_{\beta\gamma}-
\delta_{\alpha \beta } \delta_{ \gamma \delta})+\zeta \delta_{\alpha \beta } \delta_{ \gamma \delta}
\\&&
\label{eq:odd_viscous_stress}
\eta^{\text{odd}}_{\alpha\beta\gamma\delta}=\eta_o (\epsilon_{\alpha \gamma}\delta_{\beta \delta}+\epsilon_{\beta \delta} \delta_{\alpha \gamma})\,\rm{.}
\eea
\end{subequations}
From Eq.~\eqref{eq:odd_viscous_stress}, we can write the components of $\tilde{\sigma}_{\alpha\beta}$:
\bea&&
\tilde \sigma^{\text{odd}}_{\alpha\beta}=-
\eta_{o} \begin{pmatrix}
-\left( \partial_x u_y+ \partial_y u_x\right) & \partial_x u_x- \partial_y u_y \\
  \partial_x u_x- \partial_y u_y & \partial_x u_y+ \partial_y u_x
\end{pmatrix}\,\rm{,}
\label{avronxy}
\eea
which satisfies the relations $\tilde \sigma^{\text{odd}}_{xx} = - \tilde \sigma^{\text{odd}}_{yy}$ and $\tilde \sigma^{\text{odd}}_{xy} = \tilde \sigma^{\text{odd}}_{yx}$.

\section{Fourier transform}
\label{app:FourierAppendix}

In this Appendix, we define the Fourier transform utilized to analytically solve the linearized hydrodynamic equations~\eqref{raph1}. Specifically, we have applied the Fourier transform to density and velocity fields, switching from the real space representation to inverse space, described by the wavevector $\mathbf{q}$.
The spatial Fourier transform of a function $\ff(\rr, t)$ is denoted by $\hat{\ff} (\bq, t)$ and is defined as
\begin{equation}
\label{eq:fourier_transform}
\hat{\ff} (\bq, t) = \int_{\mathbb{R}^{2}} d^{2}\rr \,\ff(\rr, t) e^{-i\bq \cdot \rr}\,\rm{.}
\end{equation}
Consequently, one can reconstruct $\ff(\rr, t)$ using the inverse Fourier transform
\begin{equation}
\label{eq:invfourier_transform}
\ff (\rr, t) = \int_{\mathbb{R}^2} \dfrac{d^2\bq}{(2\pi)^{2}} \,\hat{\ff}(\bq, t) e^{i\bq \cdot \rr}\,\rm{.}
\end{equation}

\section{Effect of odd viscosity on the diffusion properties}\label{app:ficklaw_oddviscosity}

In this Appendix, we show the effect of viscous terms on the long-time diffusive properties of a chiral active fluid, described by the hydrodynamic theory given by Eqs.~\eqref{NSeq} and~\eqref{eq:continuity}. 
In a standard fluid, the dynamics of the transverse momentum are decoupled from those of the longitudinal momentum. This is not the case in the present model. We start by linearizing the continuity and velocity equations (Eqs.~\eqref{continuity1} and~\eqref{velocita}) around the bulk density $n_b$ and zero velocity $\bu={\bf0}$,
\beq
\frac{\partial }{\partial t} \hat n(\bq,t)+i q n_b \hat u_L(\bq,t)=0 \, \rm{.}
\label{qcontinuity}
\eeq
After separating the longitudinal and tangential velocity components, $\hat u_L$ and $\hat u_T$, we find that the time derivatives of the velocities can be neglected for a long time, since only $\hat n(\bq,t)$ is conserved. This leads to
\bea&&
(\gamma+ (\nu + \nu_b) q^2) \hat u_L+\nu_o q^2 \hat u_T=i q \left( \frac{c_s^2}{n_b}+\frac{\kappa}{m} q^2\right) \hat n \\&&
(\gamma+ \nu q^2) \hat u_T-\nu_o q^2\hat  u_L=iq\frac{\chi_o}{m} \hat n \,\rm{.}
\eea
Solving for $\hat{u}_L$, we obtain
\beq
\hat u_L(\bq)=  -iq \frac{ (\gamma+ \nu q^2) \left(\frac{c_s^2}{n_b}+\frac \kappa m q^2\right)-\frac{\nu_o \chi_o}{m} q^2   }{  (\gamma+ \nu q^2)(\gamma+ (\nu+\nu_b) q^2)+\nu_o^2 q^4      }               \hat  n(\bq,t)\,\rm{,}
\label{uldiff}
\eeq
which can be substituted into Eq.~\eqref{qcontinuity}. Expanding to order $q^4$, we find
\beq
\frac{\partial }{\partial t}\hat n(\bq,t) =  -\left( \frac{c_s^2}{\gamma} q^2+ \left(\frac{\kappa n_b}{m\gamma}  -\frac{\chi_o n_b \nu_o}{m\gamma^2}   -\frac{c_s^2 (\nu + \nu_b)}{\gamma^2} \right)q^4 \right) \hat n(\bq,t) + \mathcal{O}(q^{6}) \,\rm{.}
\label{qdiffusion}
\eeq
One can also show that, in the presence of odd-parity effective forces, the system develops a transverse current $n_b u_T$:
\beq
\hat u_T(\bq)=-iq\frac{ (\gamma+ (\nu+\nu_b) q^2) \frac{\chi_o}{m}+\left( \frac{c_s^2}{n_b}+\frac{\kappa}{m} q^2\right)q^2 \nu_o   }{  (\gamma+ \nu q^2)(\gamma+ (\nu + \nu_b) q^2)+\nu_o^2 q^4      }                \hat n(\bq,t) \,\rm{,}
\eeq
which does not contribute to bulk diffusion.
Note that, when both $\chi_o=0$ and $\nu_o=0$, this component of the current vanishes, as it is not coupled to density fluctuations in this case.

\section{Details on the linear stability analysis}\label{app:DetailsLinearStability}

In this Appendix, we investigate the sign of the eigenvalues $\lambda(q)$ of the dynamical matrix:
\begin{equation}
   \mathbf  M(q)=\begin{pmatrix}
0 & -i n_b q & 0\\
-i q \Big(\frac{c_s^2}{n_b}+\frac{\kappa}{m} q^2\Big)&-\gamma-(\nu+\nu_b)q^2&-\nu_oq^2\\
- i q\frac{\chi_o}{m} &\nu_o q^2&-\gamma-\nu q^2
\end{pmatrix}\,\rm{.}
\label{eq: M hydro}
\end{equation}
The eigenvalues $\lambda(q)$ are obtained from the characteristic equation $\det\big(\lambda(q)\mathbf I - \mathbf M(q)\big)=0$ which yields:
\begin{equation}
    A_0(q)+A_1(q)\lambda+A_2(q)\lambda^2-\lambda^3=0,
    \label{eq: charact}
\end{equation}
with
\begin{equation}
\begin{aligned}
    A_0(q)&=-c_s^2\gamma q^2+\Big(\frac{\chi_o\nu_o n_b}{m}-c_s^2\nu-\dfrac{\gamma\kappa n_b}{m}\Big)q^4 - \dfrac{n_b\kappa\nu}{m}q^6\,\rm{,}\\
    A_1(q)&=-\gamma^2-\Big(c_s^2+\gamma(2\nu+\nu_b)\Big)q^2-\Big(\nu_o^2+\nu^2+\nu\nu_b+n_b\kappa/m\Big)q^4\,\rm{,}\\
    A_2(q)&=-(2\nu+\nu_b)q^2-2\gamma\,\rm{.}
\end{aligned}
\end{equation}
This equation is cubic in $\lambda$ and, although exact closed-form solutions exist, they are cumbersome and offer limited physical insight. Instead, we analyze the system in various limits and approximations to elucidate its behavior.

\subsection{Stability of the system for vanishing odd viscosity}

For vanishing odd viscosity $\nu_o=0$, the characteristic polynomial Eq.~\eqref{eq: charact} becomes independent of $\chi_o$ and therefore reduces to the one of a non-chiral damped fluid, which is stable. Therefore, a fluid with chiral torque alone but no odd viscosity remains linearly stable.

\subsection{Small-\texorpdfstring{$q$}{q} expansion of the eigenvalues}

We now obtain approximate expressions for the eigenvalues in the low $q$ limit when $\nu_o\neq 0$. Either by solving the cubic characteristic equation for $M$ exactly and expanding it to order $q^2$, or by constructing a perturbative expansion of the eigenvalues in powers of $q$, or even by looking for solutions in $\lambda = -\gamma + \lambda_{\pm}q^2$ and $\lambda = \lambda_0q^2$, we obtain:
\begin{subequations}
\bea
\lambda_D(q) &=& -\frac{c_s^2}{\gamma}q^2+\mathcal O(q^4),\\
\lambda_{T}(q) &=& -\gamma-\left(\nu+\frac{\nu_b}{2}-\frac{c_s^2}{2\gamma}
+\frac{1}{2\gamma}\sqrt{(c_s^2-\gamma\nu_b)^2-4\nu_o\left(\gamma^2\nu_o+\frac{\chi_o\gamma n_b}{m}\right)}\right)q^2+\mathcal O(q^4),\\
\lambda_{L}(q) &=& -\gamma-\left(\nu+\frac{\nu_b}{2}-\frac{c_s^2}{2\gamma}
-\frac{1}{2\gamma}\sqrt{(c_s^2-\gamma\nu_b)^2-4\nu_o\left(\gamma^2\nu_o+\frac{\chi_o\gamma n_b}{m}\right)}\right)q^2+\mathcal O(q^4).
\eea
\end{subequations}
Expanding these eigenvalues in small $\epsilon$, with:
\begin{equation}
    \epsilon= \frac{4\nu_o(m\gamma^2\nu_o+\chi_o\gamma n_b)}{m(c_s^2-\gamma\nu_b)^2}\,\rm{,}
\end{equation}
we recover Eqs.~\eqref{eigenvalue_expanded} of the main text.

\subsection{Large \texorpdfstring{$q$}{q} expansion of the eigenvalues}

We set $\kappa=0$ from now on. In the limit $q\to\infty$, we numerically found that two eigenvalues diverge; therefore, we can try a solution of the form:
\begin{equation}
    \lambda(q\to\infty)=\lambda_\pm q^2 + \mathcal O(1)\,\rm{,}
    \label{eq: ansatz}
\end{equation}
where we anticipate that there would be two solutions $\lambda_\pm$. By replacing Eq.~\eqref{eq: ansatz} into Eq.~\eqref{eq: charact}, we find:
\begin{equation}
    \lambda_\pm^2+(2\nu+\nu_b)\lambda+(\nu_o^2+\nu^2 + \nu\nu_b)=\mathcal O(q^{-2})\,\rm{.}
    \label{eq: quadratic kpm}
\end{equation}
The terms $\lambda_{\pm}$ are therefore solutions to a quadratic equation. Solving Eq.~\eqref{eq: quadratic kpm} leads to Eqs.~\eqref{eq:lamnda1} and~\eqref{eq:lamnda2} in the main text.
To obtain the third equation, we try the Ansatz:
\begin{equation}
    \lambda(q\to\infty)=\lambda_0\,\rm{.}
    \label{eq: ansatz2}
\end{equation}
Replacing Eq.~\eqref{eq: ansatz2} into Eq.~\eqref{eq: charact} leads to:
\begin{equation}
    \Big(\frac{\chi_o\nu_o n_b}{m}-c_s^2\nu\Big)-\Big(\nu_o^2+\nu^2+\nu\nu_b\Big)\lambda_0=\mathcal O(q^{-2})\,\rm{,}
\end{equation}
which shows Eq.~\eqref{eq:lamnda3} of the main text.

\subsection{Unstable wavevector \texorpdfstring{$q_c$}{qc}}

To find the wavevector at which at least one eigenvalue changes sign, we can look for a solution with $\lambda(q_c)=0$. From Eq.~\eqref{eq: charact}, we obtain
\begin{equation}
    -mc_s^2\gamma +\Big(\chi_o\nu_o n_b-mc_s^2\nu\Big)q_c^2=0\,\rm{.}
    \label{eq: qc instability}
\end{equation}
This is Eq.~\eqref{eq: qc main text} of the main text.

\subsection{Effect of \texorpdfstring{$\kappa$}{k} on the eigenvalues}

In most of our asymptotic calculations, we set $\kappa = 0$, thereby neglecting the contribution of the Korteweg stress tensor. Physically, a finite $\kappa > 0$ penalizes sharp density gradients and acts as an ultraviolet regularization of the hydrodynamic theory: it suppresses short-wavelength fluctuations and stabilizes the large-$q$ regime. In particular, for $\kappa > 0$, the real parts of all three eigenvalues are unconditionally negative in the limit $q \to \infty$, $\mathrm{Re}[\lambda](q \to \infty) \to -\infty$. Nevertheless, provided that $\kappa$ is not unphysically large, there remains a finite interval of wavevectors for which at least one eigenvalue becomes positive, $\mathrm{Re}[\lambda](q \in [q_c^{(-)}, q_c^{(+)}]) > 0$. In the limit $\kappa \to 0$, we recover $q_c^{(-)} = q_c$ and $q_c^{(+)} \to \infty$. The critical wavevectors $q_c^{(\pm)}$ are obtained by solving the quadratic equation resulting from the condition $\lambda(q_c^{(\pm)}) = 0$.

Accordingly, rather than setting $\kappa = 0$ and requiring $q_c \lesssim 1$ for an instability to exist, one may keep $\kappa$ finite and rely on it to stabilize the large-wavevector sector. In principle, this removes the need to introduce an ad hoc cutoff. In practice, however, it complicates the analysis without providing additional physical insight and, in the absence of large density gradients (e.g., near interfaces), it typically contributes only at wavevectors beyond the domain of applicability of the hydrodynamic equations.

\section{Navier-Stokes equation in polar coordinates}
\label{polarcoordinates}

The cavity-fluid problem requires the solution of the Navier-Stokes equation in polar coordinates in the presence of friction.
In order to investigate the phase characterized by nearly circular cavities, we consider a single circular cavity and adopt polar coordinates, assuming a steady ($\frac{\partial }{\partial t}=0$), axisymmetric ($\frac{\partial }{\partial \theta}=0$) state.
The continuity and the Navier-Stokes equation in this representation read:
\bea&&
\label{eq:linear_density_eq}
\dfrac{1}{r}\partial_r (r n u_r)  =0
\\&&
\label{eq:linear_nsx_eq}
u_r\partial_r u_r-\dfrac{u_\theta^2}{r}=
-\dfrac{1}{m n} \partial_r\left(P-\kappa n(r)\nabla^2 n(r)\right)+\dfrac{F_r}{m}-\gamma  u_r
+\nu_o \Bigl( \partial^2_r u_\theta +\dfrac{1}{r}\partial_r u_\theta
 -\dfrac{1}{r^2}u_\theta    \Bigr)
\nonumber
\\&&
 + \nu \Bigl( \partial^2_r u_r +\dfrac{1}{r}\partial_r u_r -\dfrac{1}{r^2}u_r    \Bigr)
\\&&
\label{eq:linear_nsy_eq}
u_r\partial_r u_\theta +\dfrac{u_\theta u_r}{r}=
\dfrac{F_\theta}{m}-\gamma u_\theta+\nu \Bigl( \partial^2_r u_\theta +\dfrac{1}{r}\partial_r u_\theta
 -\dfrac{1}{r^2}u_\theta    \Bigr)
 \nonumber
 \\&& -\nu_o \Bigl( \partial^2_r u_r +\dfrac{1}{r}\partial_r u_r-\dfrac{1}{r^2}u_r    \Bigr)\,\rm{.}
\eea
If we further assume no radial flux ($n  u_r=0$) and a vanishing external radial field $F_{r}$, Eq.~\eqref{eq:linear_density_eq} is trivial and the hydrodynamic solutions $(n, u_{\theta})$ satisfies Eqs.~\eqref{eq:linear_nsx_eq}-~\eqref{eq:linear_nsy_eq}, where the tangential component $F_\theta$ represents the tangential force:
$
F_\theta=
 |{\bf f}^{\perp}(\rr)|=
|\chi_o\,  (\hat{\bf z} \times \bfnabla n(\rr))|$. Under these approximations, we can rewrite the hydrodynamical system as 
\bea&&
\label{eq:linear_nsx_eq2}
-\dfrac{u_\theta^2}{r}=
-\dfrac{1}{m n} \partial_r\left(P-\kappa n(r)\nabla^2 n(r)\right)
+\nu_o \Bigl( \partial^2_r u_\theta +\dfrac{1}{r}\partial_r u_\theta
 -\dfrac{1}{r^2}u_\theta    \Bigr)
\\&&
\label{eq:linear_nsy_eq2}
0=\dfrac{F_\theta}{m}-\gamma u_\theta+\nu \Bigl( \partial^2_r u_\theta +\dfrac{1}{r}\partial_r u_\theta
 -\dfrac{1}{r^2}u_\theta    \Bigr) \,\rm{.}
\eea
Using the polar form of the vorticity $\omega_z(r)=\frac{1}{r}\partial_r (r u_\theta(r))$ and the definition of odd pressure $P_o=\eta_o \omega_z$, we can write
\bea&&
\dfrac{u_\theta^2}{r}=\dfrac{1}{m n} \dfrac{\partial }{\partial r} (P-P_o)+\dfrac{\kappa}{m} \dfrac{\partial}{\partial r}  \nabla^2 n(r)
\nonumber
\\&&
\dfrac{F_\theta}{m}-\gamma u_\theta+\dfrac{\nu}{\nu^{o}} \dfrac{1}{m n} \dfrac{\partial }{\partial r} P_o =0 \,\rm{.}
\eea
For the sake of simplicity, we shall neglect the standard and the odd viscosities when solving the equations for the vortices.
With this approximation, we do not see the dependence upon $\nu_o$ and $\nu$ on the linear instability, since the viscosity coefficients will not appear hereafter. Therefore, the theory cannot describe the initial formation of bubbles. It can only describe their final state, which is described by the following radial and tangential force balances
\bea&&
-\dfrac{u_\theta^2}{r}=
-\dfrac{1}{mn} \partial_r P +\kappa \Bigl( \dfrac{\partial^3 n }{\partial r^3}  +\frac{1}{r}  \frac{\partial^2 n}{\partial r^2}  
-  \frac{1}{r^2}  \frac{\partial n}{\partial r}  \Bigr)
\label{Radialbalance2} \\&&
 \chi_o \partial_r n(r)=m\gamma u_\theta \,.
 \label{Tangentialbalance2}
\eea
Eliminating $u_\theta(r)$ in  favor of $n(r)$ and using Eq.~\eqref{Radialbalance2} we obtain Eq.~\eqref{equazioneradialea}.
In summary, by neglecting viscosities -- i.e., in the inviscid limit we have to solve an Euler hydrodynamic equation with a friction term -- we get a closed non-linear equation for the density.

\section*{References}

\bibliographystyle{apsrev4-1}

\bibliography{oddhydrodynamicsbibliography.bib}

\end{document}